\pdfoutput=1
\documentclass{article}

\usepackage[preprint]{neurips_2026_vericode_competition}

\usepackage[utf8]{inputenc} %
\usepackage[T1]{fontenc}    %
\usepackage{amsmath}        %
\usepackage{hyperref}       %
\usepackage{url}            %
\usepackage{booktabs}       %
\usepackage{amsfonts}       %
\usepackage{nicefrac}       %
\usepackage{microtype}      %
\usepackage{xcolor}         %

\usepackage{graphicx}       %
\usepackage{tikz}           %
\usetikzlibrary{arrows.meta,positioning,fit,calc}
\usepackage{multirow}
\usepackage{longtable}      %
\usepackage{listings}       %

\definecolor{vxBlue}{HTML}{0072B2}
\definecolor{vxOrange}{HTML}{E69F00}
\definecolor{vxSky}{HTML}{56B4E9}
\definecolor{vxGreen}{HTML}{009E73}
\definecolor{vxVermillion}{HTML}{D55E00}
\definecolor{vxPurple}{HTML}{CC79A7}
\definecolor{vxGray}{HTML}{6E6E6E}
\colorlet{vxOpen}{vxBlue}
\colorlet{vxClosed}{vxOrange}
\colorlet{vxSeed}{vxGray}

\lstdefinelanguage{Lean4}{
  morekeywords={theorem,lemma,by,fun,with,at,only,generalizing,in,have,show,from,let,if,then,else,match,do},
  morekeywords=[2]{induction,fun_induction,cases,simp,simp_all,grind,split,rfl,exact,intro,all_goals,specialize,rename_i,omega,decide,unhygienic},
  sensitive=true,
  morecomment=[l]{--},
  morecomment=[s]{/-}{-/},
  alsoletter={_!?.'},
}
\makeatletter
\newcommand{\fact}[1]{%
  \@ifundefined{#1}%
    {\textcolor{red}{[#1?]}\typeout{VX-MISSING-FACT: #1}}%
    {\csname #1\endcsname}}
\makeatother
\makeatletter
\newcommand{\vxcore}[1]{\@ifundefined{OpenCoreTrack}{\fact{ClosedCore#1}}{\fact{OpenCore#1}}}
\newcommand{\vxifopen}[2]{\@ifundefined{OpenCoreTrack}{#2}{#1}}
\makeatother

\newcommand{\vxbodyend}{\label{vx:bodyend}}

\newcommand{\vxreferences}{%
  \section*{References}\label{vx:references}%
  {\small
   \renewcommand{\bibsection}{}%
   \bibliographystyle{plainnat}%
   \bibliography{common/references}}%
}

\usetikzlibrary{plotmarks,shapes.geometric,decorations.pathreplacing}
\hypersetup{pdftitle={Compile-Gated Lean Proof Refactoring with Enterprise Agent Practices: A Model per Step under a US\$3 Cap, Followed by Programmatic Rewrite Rules}, pdfauthor={Bobber Cheng}}   %

\newcommand{\ClosedFigScoreDiffMax}{0.10}
\newcommand{\ClosedFigScoreDiffRows}{13}

\newcommand{\ClosedPerProblemRows}{%
\midrule\multicolumn{9}{l}{\textbf{PutnamBench}}\\
\texttt{putnam\_\allowbreak{}1964\_\allowbreak{}a4}$^\dagger$ & 1.239 & 1.516 & 0.097 & 2.852 & 2.949 & 97.43 & 94.5 & 97.8\\
\texttt{putnam\_\allowbreak{}1995\_\allowbreak{}a3}$^\dagger$ & 1.225 & 1.530 & 0.094 & 2.848 & 2.942 & 97.05 & 97.6 & 93.6\\
\texttt{putnam\_\allowbreak{}1988\_\allowbreak{}a2} & 1.066 & 1.695 & 0.070 & 2.831 & 2.901 & 96.85 & 94.8 & 95.7\\
\texttt{putnam\_\allowbreak{}2001\_\allowbreak{}b1} & 1.184 & 1.525 & 0.117 & 2.827 & 2.944 & 96.60 & 94.1 & 95.7\\
\texttt{putnam\_\allowbreak{}1964\_\allowbreak{}b2}$^\dagger$ & 1.153 & 1.670 & 0.051 & 2.874 & 2.925 & 99.29 & 98.6 & 99.3\\
\texttt{putnam\_\allowbreak{}2007\_\allowbreak{}a1} & 1.183 & 1.517 & 0.096 & 2.795 & 2.892 & 98.81 & 98.5 & 97.9\\
\texttt{putnam\_\allowbreak{}1970\_\allowbreak{}b5} & 1.196 & 1.479 & 0.133 & 2.808 & 2.942 & 98.28 & 96.1 & 98.7\\
\texttt{putnam\_\allowbreak{}1968\_\allowbreak{}a1} & 1.167 & 1.468 & 0.145 & 2.780 & 2.925 & 95.77 & 93.1 & 94.2\\
\texttt{putnam\_\allowbreak{}1991\_\allowbreak{}b1} & 1.101 & 1.483 & 0.191 & 2.775 & 2.966 & 97.26 & 93.9 & 97.9\\
\texttt{putnam\_\allowbreak{}2019\_\allowbreak{}a1} & 1.056 & 1.673 & 0.077 & 2.806 & 2.883 & 98.09 & 94.6 & 99.6\\
\midrule\multicolumn{9}{l}{\textbf{ArkLib}}\\
\texttt{interleaved\_\allowbreak{}affine\_\allowbreak{}gaps\_\allowbreak{}imply\_\allowbreak{}tensor\_\allowbreak{}gaps}$^\dagger$ & 1.053 & 1.538 & 0.152 & 2.743 & 2.896 & 93.16 & 93.3 & 86.1\\
\texttt{Binius.\allowbreak{}BinaryBasefold.\allowbreak{}fold\_\allowbreak{}advances\_\allowbreak{}evaluation\_\allowbreak{}poly}$^\dagger$ & 1.095 & 1.624 & 0.084 & 2.803 & 2.887 & 94.65 & 94.8 & 89.2\\
\texttt{Binius.\allowbreak{}BinaryBasefold.\allowbreak{}fiberwise\_\allowbreak{}dist\_\allowbreak{}lt\_\allowbreak{}imp\_\allowbreak{}dist\_\allowbreak{}lt\_\allowbreak{}unique\_\allowbreak{}decoding\_\allowbreak{}radius}$^\dagger$ & 1.083 & 1.554 & 0.123 & 2.760 & 2.882 & 91.91 & 87.7 & 88.0\\
\texttt{Combine.\allowbreak{}master\_\allowbreak{}lemma} & 1.144 & 1.547 & 0.118 & 2.808 & 2.926 & 92.09 & 78.9 & 97.4\\
\texttt{JohnsonBound.\allowbreak{}johnson\_\allowbreak{}bound\_\allowbreak{}alphabet\_\allowbreak{}free} & 1.064 & 1.569 & 0.136 & 2.769 & 2.906 & 84.69 & 69.3 & 84.8\\
\texttt{ProximityToR\allowbreak{}S.\allowbreak{}dir\_\allowbreak{}close\_\allowbreak{}of\_\allowbreak{}many\_\allowbreak{}close\_\allowbreak{}pts} & 1.183 & 1.526 & 0.118 & 2.827 & 2.945 & 96.28 & 92.6 & 96.3\\
\texttt{ProximityGap.\allowbreak{}RS\_\allowbreak{}exists\_\allowbreak{}nonzero\_\allowbreak{}kernelVec\_\allowbreak{}of\_\allowbreak{}det\_\allowbreak{}eq\_\allowbreak{}zero\_\allowbreak{}natDegree\_\allowbreak{}le\_\allowbreak{}one} & 1.079 & 1.492 & 0.170 & 2.741 & 2.910 & 93.83 & 83.6 & 97.8\\
\texttt{ProximityGap.\allowbreak{}RS\_\allowbreak{}jointAgreement\_\allowbreak{}of\_\allowbreak{}goodCoeffs\_\allowbreak{}card\_\allowbreak{}gt} & 1.059 & 1.503 & 0.187 & 2.749 & 2.936 & 90.58 & 83.1 & 88.7\\
\texttt{Sumcheck.\allowbreak{}Spec.\allowbreak{}SingleRound.\allowbreak{}Simple.\allowbreak{}reduction\_\allowbreak{}perfectCompleteness} & 1.075 & 1.526 & 0.178 & 2.779 & 2.957 & 90.63 & 80.8 & 91.1\\
\texttt{ps\_\allowbreak{}coprime\_\allowbreak{}case\_\allowbreak{}constant} & 1.056 & 1.521 & 0.175 & 2.753 & 2.928 & 87.45 & 68.1 & 94.3\\
\midrule\multicolumn{9}{l}{\textbf{Strata}}\\
\texttt{Core.\allowbreak{}InitsUpdates\allowbreak{}Comm}$^\dagger$ & 1.143 & 1.646 & 0.054 & 2.843 & 2.896 & 87.72 & 68.3 & 94.9\\
\texttt{Lambda.\allowbreak{}denote\_\allowbreak{}suffix\_\allowbreak{}irrel} & 1.101 & 1.525 & 0.138 & 2.764 & 2.902 & 77.05 & 59.3 & 71.9\\
\texttt{Lambda.\allowbreak{}denote\_\allowbreak{}Denotes} & 1.195 & 1.479 & 0.133 & 2.806 & 2.939 & 88.72 & 81.1 & 85.1\\
\texttt{Lambda.\allowbreak{}callOfLFunc\_\allowbreak{}eraseMetadata\_\allowbreak{}congr} & 1.102 & 1.590 & 0.104 & 2.796 & 2.900 & 92.77 & 85.6 & 92.7\\
\texttt{Lambda.\allowbreak{}liftBVars\_\allowbreak{}typeCheck} & 1.098 & 1.656 & 0.070 & 2.825 & 2.895 & 92.86 & 88.0 & 90.6\\
\texttt{Lambda.\allowbreak{}liftBVars\_\allowbreak{}denote} & 1.089 & 1.506 & 0.180 & 2.775 & 2.955 & 77.41 & 67.1 & 65.2\\
\texttt{Lambda.\allowbreak{}LExpr.\allowbreak{}typeCheck\_\allowbreak{}to\_\allowbreak{}HasTypeA} & 1.098 & 1.577 & 0.103 & 2.779 & 2.882 & 89.36 & 85.8 & 82.3\\
\texttt{Lambda.\allowbreak{}callOfLFunc\_\allowbreak{}constr\_\allowbreak{}disjoint\_\allowbreak{}denote} & 1.093 & 1.476 & 0.166 & 2.735 & 2.900 & 75.84 & 59.2 & 68.3\\
\texttt{Lambda.\allowbreak{}denote\_\allowbreak{}replaceMetadata} & 1.065 & 1.512 & 0.163 & 2.740 & 2.904 & 95.13 & 94.3 & 91.1\\
\texttt{Lambda.\allowbreak{}denote\_\allowbreak{}ext} & 1.065 & 1.487 & 0.162 & 2.715 & 2.877 & 77.63 & 80.2 & 53.0\\
\midrule\multicolumn{9}{l}{\textbf{LeanPool}}\\
\texttt{LeanPool.\allowbreak{}Erdos132Convex\allowbreak{}K3.\allowbreak{}erlv\_\allowbreak{}case12\_\allowbreak{}other\_\allowbreak{}d1\_\allowbreak{}bone\_\allowbreak{}impossible} & 1.069 & 1.494 & 0.176 & 2.738 & 2.914 & 94.45 & 86.6 & 96.8\\
\texttt{CoxeterSystem.\allowbreak{}matsumoto\_\allowbreak{}reduced\_\allowbreak{}aux} & 1.052 & 1.574 & 0.142 & 2.767 & 2.909 & 86.63 & 78.2 & 81.7\\
\texttt{Vlasov.\allowbreak{}transportedIntegral\_\allowbreak{}continuousOn} & 1.101 & 1.569 & 0.095 & 2.765 & 2.861 & 85.77 & 81.1 & 76.2\\
\texttt{ZetaZeros.\allowbreak{}step1\_\allowbreak{}double} & 1.172 & 1.567 & 0.100 & 2.839 & 2.939 & 92.93 & 91.0 & 87.8\\
\texttt{LocalComplex\allowbreak{}Geometry.\allowbreak{}holomorphic\_\allowbreak{}constant\_\allowbreak{}rank} & 1.081 & 1.530 & 0.155 & 2.765 & 2.921 & 71.14 & 68.1 & 45.4\\
\texttt{LeanEval.\allowbreak{}Topology.\allowbreak{}Classification\allowbreak{}OfSurfaces.\allowbreak{}Moise.\allowbreak{}PartialTriangulation.\allowbreak{}exists\_\allowbreak{}chartMatching\allowbreak{}ControlOn\_\allowbreak{}of\_\allowbreak{}metricSpace} & 1.056 & 1.553 & 0.157 & 2.766 & 2.924 & 83.75 & 73.4 & 77.9\\
\texttt{LeanEval.\allowbreak{}Topology.\allowbreak{}Classification\allowbreak{}OfSurfaces.\allowbreak{}Moise.\allowbreak{}PolygonalCircle.\allowbreak{}insertZero\_\allowbreak{}nonadjacent\_\allowbreak{}disjoint} & 1.181 & 1.508 & 0.123 & 2.812 & 2.936 & 86.92 & 77.3 & 83.4\\
\texttt{IsoGraph.\allowbreak{}Canon.\allowbreak{}dfsNode\_\allowbreak{}dom} & 1.065 & 1.498 & 0.156 & 2.719 & 2.875 & 69.48 & 50.7 & 57.8\\
\texttt{FirstOrder.\allowbreak{}Language.\allowbreak{}schemaTruthStage\_\allowbreak{}imp} & 1.062 & 1.571 & 0.166 & 2.799 & 2.966 & 89.13 & 70.4 & 97.0\\
\texttt{RS.\allowbreak{}cechCount} & 1.084 & 1.496 & 0.166 & 2.747 & 2.913 & 91.58 & 78.7 & 96.0\\
\midrule\multicolumn{9}{l}{\textbf{CSLib}}\\
\texttt{Cslib.\allowbreak{}LambdaCalculus.\allowbreak{}LocallyNameless.\allowbreak{}Fsub.\allowbreak{}Typing.\allowbreak{}progress}$^\dagger$ & 1.079 & 1.546 & 0.140 & 2.765 & 2.906 & 91.38 & 83.1 & 91.1\\
\texttt{Cslib.\allowbreak{}SKI.\allowbreak{}parallelReduction\_\allowbreak{}diamond}$^\dagger$ & 1.064 & 1.492 & 0.186 & 2.742 & 2.928 & 77.93 & 62.9 & 70.9\\
\texttt{Cslib.\allowbreak{}CCS.\allowbreak{}bisimilarity\_\allowbreak{}congr\_\allowbreak{}choice}$^\dagger$ & 1.059 & 1.570 & 0.170 & 2.799 & 2.969 & 88.54 & 72.9 & 92.8\\
\texttt{Cslib.\allowbreak{}LambdaCalculus.\allowbreak{}LocallyNameless.\allowbreak{}Fsub.\allowbreak{}Typing.\allowbreak{}preservation} & 1.101 & 1.481 & 0.191 & 2.773 & 2.964 & 83.63 & 66.0 & 85.0\\
\texttt{Cslib.\allowbreak{}FLP.\allowbreak{}PseudoConsensus.\allowbreak{}of\_\allowbreak{}consensus} & 1.106 & 1.518 & 0.130 & 2.754 & 2.884 & 78.37 & 43.2 & 91.9\\
\texttt{Cslib.\allowbreak{}Automata.\allowbreak{}NA.\allowbreak{}Buchi.\allowbreak{}buchiFamily\_\allowbreak{}saturation} & 1.058 & 1.504 & 0.183 & 2.745 & 2.928 & 80.93 & 50.9 & 91.9\\
\texttt{Cslib.\allowbreak{}FLP.\allowbreak{}OnePseudoConsensus.\allowbreak{}nonUniform\_\allowbreak{}step} & 1.188 & 1.509 & 0.113 & 2.811 & 2.924 & 91.94 & 78.3 & 97.6\\
\texttt{Cslib.\allowbreak{}LTS.\allowbreak{}OmegaExecution.\allowbreak{}flatten\_\allowbreak{}execution} & 1.176 & 1.510 & 0.121 & 2.807 & 2.928 & 77.91 & 61.9 & 71.9\\
\texttt{Cslib.\allowbreak{}FLP.\allowbreak{}CanReachVia.\allowbreak{}subset\_\allowbreak{}inp} & 1.150 & 1.577 & 0.100 & 2.827 & 2.927 & 86.73 & 64.3 & 95.9\\
\texttt{Cslib.\allowbreak{}LambdaCalculus.\allowbreak{}LocallyNameless.\allowbreak{}Untyped.\allowbreak{}Term.\allowbreak{}parallel\_\allowbreak{}diamond} & 1.149 & 1.494 & 0.149 & 2.791 & 2.940 & 82.83 & 59.3 & 89.1\\
\midrule\multicolumn{9}{l}{\textbf{PhysLib}}\\
\texttt{FTheory.\allowbreak{}SU5.\allowbreak{}ChargeSpectrum.\allowbreak{}containsPheno\allowbreak{}CompletionsOf\allowbreak{}MinimallyAllows\_\allowbreak{}viableCompletions} & 1.185 & 0.609 & 0.019 & 1.813 & 1.832 & 33.33 & 0.0 & 0.0\\
\texttt{FTheory.\allowbreak{}SU5.\allowbreak{}ChargeSpectrum.\allowbreak{}isPhenoClosed\allowbreak{}Q5\_\allowbreak{}viableCharges} & 1.163 & 1.661 & 0.050 & 2.874 & 2.924 & 49.99 & 50.0 & 0.0\\
\texttt{FTheory.\allowbreak{}SU5.\allowbreak{}ChargeSpectrum.\allowbreak{}isPhenoClosed\allowbreak{}Q10\_\allowbreak{}viableCharges} & 1.166 & 1.660 & 0.045 & 2.871 & 2.917 & 49.99 & 50.0 & 0.0\\
\texttt{PauliMatrix.\allowbreak{}pauliContrDown\_\allowbreak{}mul\_\allowbreak{}pauliContr\_\allowbreak{}mul\_\allowbreak{}pauliContrDown} & 1.117 & 1.647 & 0.072 & 2.837 & 2.909 & 76.95 & 75.1 & 55.7\\
\texttt{PauliMatrix.\allowbreak{}pauliContr\_\allowbreak{}mul\_\allowbreak{}pauliContrDown\_\allowbreak{}mul\_\allowbreak{}pauliContr} & 1.132 & 1.623 & 0.083 & 2.838 & 2.921 & 73.10 & 69.4 & 49.9\\
\texttt{Electromagnetism.\allowbreak{}Electromagnetic\allowbreak{}Potential.\allowbreak{}time\_\allowbreak{}deriv\_\allowbreak{}time\_\allowbreak{}deriv\_\allowbreak{}electricField\_\allowbreak{}of\_\allowbreak{}isExtrema}$^\dagger$ & 1.063 & 1.511 & 0.190 & 2.764 & 2.954 & 87.77 & 80.5 & 82.8\\
\texttt{FieldSpecification.\allowbreak{}WickAlgebra.\allowbreak{}\ensuremath{\iota}\_\allowbreak{}timeOrderF\_\allowbreak{}superCommute\allowbreak{}F\_\allowbreak{}eq\_\allowbreak{}time}$^\dagger$ & 1.166 & 1.513 & 0.106 & 2.785 & 2.891 & 87.26 & 78.4 & 83.4\\
\texttt{fundamental\_\allowbreak{}theorem\_\allowbreak{}of\_\allowbreak{}variational\_\allowbreak{}calculus'}$^\dagger$ & 1.171 & 1.466 & 0.122 & 2.759 & 2.881 & 92.62 & 89.0 & 88.9\\
\texttt{HermitianMat.\allowbreak{}inv\_\allowbreak{}convex} & 1.151 & 1.565 & 0.103 & 2.818 & 2.921 & 96.55 & 91.9 & 97.7\\
\texttt{cauchyBinet} & 1.077 & 1.509 & 0.142 & 2.728 & 2.871 & 93.83 & 83.4 & 98.1\\
\midrule
Total & 66.95 & 91.73 & 7.64 & 166.33 & 173.97 & & & \\
Mean & 1.116 & 1.529 & 0.127 & 2.772 & 2.899 & 86.309 & 77.05 & 81.89\\
Maximum & 1.239 & 1.695 & 0.191 & 2.874 & 2.969 & & & \\
}

\newcommand{\ClosedExpLogRows}{%
\midrule \multicolumn{4}{@{}l}{\textit{Before the release: engine and first capped pipeline}}\\*
  10-06 & warm-up & vendor agent CLIs or our own harness for hosted models & our own harness (one tool set, every call metered) \\
  10-09 & warm-up & Claude Code with Opus~5.5 and Codex with GPT-6~Astra, uncapped, resumed passes & both plateaued; their general practices fed our harness; their proofs are in no package \\
  10-10 & warm-up & free lexer-level rewrites & adopted as a \$0 normalizer inside cleanups and the end stage \\
  10-10 & warm-up & why GPT-6.1 Sol beats our best proofs on some problems & sessions fenced off from other proofs; a check that each proof starts from the reference; briefs with facts, no proofs \\
  10-11 & warm-up & reasoning effort and tool-output size against cost & Sol at medium effort with full tool output \\
  10-11 & warm-up & problem-keyed briefs; a term-mode card & card rejected; briefs later ruled out by the start-from-the-reference rule \\
  10-12 & warm-up, held-out & Claude Code with Opus~5.5 as the capped engine, metered from a subscription & superseded by the OpenRouter-only rule \\
  10-12 & held-out & uncapped vendor agents (Claude Code with Opus~5.5, then Fable~5.1) on unseen problems & a level well below the warm-up level \\
  10-13 & warm-up & Sol from the reference with general knowledge only, capped & fresh restarts dropped; the briefs' head start became the CPU auto-brief's target \\
  10-13 & warm-up & spend lost to runaway tool calls & a guard from the first round and a billing reconcile \\
  10-14 & warm-up & Gemini~4 Argon as the main model & not adopted (Table~\ref{tab:decisions}) \\
  10-14 & warm-up & a CPU auto-brief instead of problem-keyed briefs & adopted as the start of the capped pipeline \\
  10-14 & warm-up & deeper or cheaper models (Opus~5.5, DeepSeek~V4.1 Flash, Kimi~K3) on hard problems & none reached the target; Sol had the most points per dollar \\
  10-14 & warm-up, held-out & \$0 rewrite families learned from winning edits & micro-edit and heartbeat families on by default, simp merging opt-in \\
  10-14 & warm-up & a tool-less architect model writing proof skeletons & dropped (no verified iteration) \\
  10-14 & warm-up, held-out & a reviewer naming edits and a DeepSeek~V4.1 Flash executor applying them & review kept, executor dropped (it pushed problems over the cap) \\
\midrule \multicolumn{4}{@{}l}{\textit{Before the release: building the capped pipeline}}\\*
  10-12 & warm-up & our meter against the Arena's & heartbeats read by our direct counter, offset calibrated \\
  10-15 & warm-up & Argon from the reference against Sol; Kimi~K3 as the reviewer & Sol stays the main model; Kimi paused (fewest points per dollar) \\
  10-16 & warm-up, held-out & single- against multi-round review; Argon against Astra as the reviewer & multi-round Argon review (Table~\ref{tab:decisions}) \\
  10-16 & warm-up, held-out & a general \$0 heartbeat search & adopted as a free pass in each cleanup \\
  10-16 & warm-up, held-out & an adaptive router with forced re-plans against one ladder & re-plan not adopted (Table~\ref{tab:decisions}) \\
  10-16 & warm-up, held-out & syntax-only rules toward our best known proofs & the remaining gap needs a model \\
  10-20 & warm-up, held-out & is a heartbeat-token module's gain general? & templated from warm-up proofs; deleted (Section~\ref{sec:negative}) \\
  10-20 & warm-up, held-out & verified gains below the acceptance bar; leaks in model-facing text & a back-to-back end stage with a small bar; a leak scan before every test \\
  10-21 & warm-up & run-to-run spread of the frozen pipeline & adoption rule tightened to large, robust effects \\
  10-21 & held-out, warm-up & the cost of stripping warm-up knowledge from prompts & kept stripped \\
  10-25 & warm-up & splitting a problem's budget (other models, twins, routes) & fewer, larger Sol sessions win (Table~\ref{tab:decisions}) \\
  10-26 & warm-up & the OpenRouter path against the development backend at the first step & fixes to the OpenRouter path \\
  10-26 & warm-up & safety margins or the full cap; how to price Argon & the full cap, enforced by a price pin, a worst-case gate and the higher of billed and list cost \\
  10-27 & warm-up & which model for each budget step & Sol first; Terra acceptable later (Table~\ref{tab:decisions}) \\
  10-27 & warm-up & why sessions ignore suggestions in their prompt & prose knowledge closed; only executed ops (Section~\ref{sec:negative}) \\
  10-27 & warm-up & where money inside the cap is wasted & keep the ladder's top; no stop rule \\
  10-27 & warm-up, held-out & whether the review earns its place & kept (Table~\ref{tab:decisions}) \\
  10-27 & build & can a failed or refused call harm spend, the best proof or the output? & a fix list and fault-injection tests \\
\midrule \multicolumn{4}{@{}l}{\textit{Before the release: freeze and rehearsal}}\\*
  10-27 & warm-up & Sol on every step, or Sol then Terra & Sol then Terra (Table~\ref{tab:decisions}) \\
  10-28 & warm-up & Terra sessions on the room the review leaves & adopted (Table~\ref{tab:decisions}) \\
  10-28 & warm-up, held-out & in-session signals, stop rules or \$0 ops for the last gap & no lever inside the cap \\
  10-28 & warm-up & a proof reaching \texttt{sorryAx} where a lakefile turns \texttt{warn.sorry} off & \texttt{warn.sorry} forced on; an axiom audit of every final proof \\
  10-28 & warm-up & a repair turn and a stack for reviewer edits & adopted (Table~\ref{tab:decisions}) \\
  10-28 & warm-up & Sol at extra-high effort for the first step & stopped; medium kept \\
  10-28 & held-out & a per-question prompt and a fresh rewrite session against the Terra steps & stopped by its pre-registered bar \\
  10-29 & warm-up, held-out & reallocation policies inside the cap & none beat the frozen allocation (Table~\ref{tab:decisions}) \\
  10-29 & warm-up & the whole library in context instead of tools & not adopted; verified iteration is the bottleneck \\
  10-29 & warm-up, recorded run states & a \$0 polish of one other approach & below its bar; adopted by the author's decision, with a time cap \\
  10-29 & warm-up, held-out & a model rewrite into the short style the score rewards & adopted (Table~\ref{tab:decisions}) \\
  10-30 & warm-up & a blind first session (statement and facts only) & not adopted (Section~\ref{sec:negative}) \\
  10-31 & warm-up & Claude Sonnet~5.5 as the first-step model & killed (Table~\ref{tab:decisions}) \\
  10-31 & warm-up & does the first session see the brief's start? & a fix so that it does, shipped in the frozen code \\
  10-31 & warm-up & a from-reference rehearsal of the frozen pipeline & frozen and launched at the release \\
  11-01 & warm-up & \$0 levers mined from our best known proofs & stopped; no lever shipped \\
\midrule \multicolumn{4}{@{}l}{\textit{After the release, on the final benchmark (in sample)}}\\*
  11-01 & final & run~1 of the frozen code & kept as a probe; its defects became run~2's fixes \\
  11-01 & final & fixes from run~1's records & shipped in run~2 (Appendix~\ref{app:disclosures}) \\
  11-01 & final & does each problem learn only from itself? & findings disclosed (Section~\ref{sec:lessons}) \\
  11-01 & final & uncapped closed-model research on problems at a structural ceiling & none of their proofs in any upload; general rules learned for the \$0 end passes \\
  11-02 & final & the code archive from its README on a clean checkout & fixed and kept as a rule for every export \\
  11-02 & final & Opus~5.5 sessions at a stall & no-go \\
  11-02 & final & paid bets against a control & dropped (Table~\ref{tab:decisions}) \\
  11-02 & final & a \$0 replay of the end passes on run~1's final proofs & shipped (Table~\ref{tab:decisions}) \\
  11-02 & final & run~2 (fixes and end passes) & within run-to-run noise of run~1 \\
  11-02 & final & run~3 on a second host & the better whole run; uploaded after the post-step \\
  11-03 & final & a fourth run with changed code & not run \\
  11-03 & final & do long sessions drift (repeat themselves, return to dead ends)? & drift not established \\
  11-03 & final & replacing heavy kernel verification & research only; fed post-step ops \\
  11-03 & final & a \$0 post-step over a finished run & adopted; reruns not byte-identical \\
  11-04 & final & keeping the kernel-check skip & withdrawn \\
  11-04 & final & post-step version~4, two passes & the final upload \\
}

\newcommand{\ClosedOpsRows}{%
  \texttt{implied} & 1 & a \texttt{decide}-family closer decides the hypothesis of a library implication an environment query finds & \fact{ClosedPostOpImpliedSteps} & \fact{ClosedPostOpImpliedRows} & \fact{ClosedPostOpImpliedGainMean} \\
  \texttt{closerprep} & 1 & \texttt{conv => reduce} before a \texttt{decide}-family closer, with a minimal drop of the simp list & \fact{ClosedPostOpCloserprepSteps} & \fact{ClosedPostOpCloserprepRows} & \fact{ClosedPostOpCloserprepGainMean} \\
  \texttt{inlineeq} & 1 & an equation proved inline in a simp list, rewritten from its own hint & \fact{ClosedPostOpInlineeqSteps} & \fact{ClosedPostOpInlineeqRows} & \fact{ClosedPostOpInlineeqGainMean} \\
  \texttt{hintgrind} & 2 & a tactic span replaced by the closer Lean prints for its goal (\texttt{grind?}, \texttt{simp?}) & \fact{ClosedPostOpHintgrindSteps} & \fact{ClosedPostOpHintgrindRows} & \fact{ClosedPostOpHintgrindGainMean} \\
  \texttt{projpath} & 2 & a destructuring pattern becomes one name and projection paths (\texttt{t.2.1}) & \fact{ClosedPostOpProjpathSteps} & \fact{ClosedPostOpProjpathRows} & \fact{ClosedPostOpProjpathGainMean} \\
  \texttt{lexsurf} & 2 & lexer-cheap surface rules (leading-dot constructors, inferable ascriptions dropped) & \fact{ClosedPostOpLexsurfSteps} & \fact{ClosedPostOpLexsurfRows} & \fact{ClosedPostOpLexsurfGainMean} \\
  \texttt{jointres} & 2 & the end-stage generators' token-saving edits, tried together; Lean's errors drop the ones that break & \fact{ClosedPostOpJointresSteps} & \fact{ClosedPostOpJointresRows} & \fact{ClosedPostOpJointresGainMean} \\
  \texttt{t4ind} & 2 & named \texttt{induction}/\texttt{cases} alternatives become bullets with Lean's names & \fact{ClosedPostOpTFourIndSteps} & \fact{ClosedPostOpTFourIndRows} & \fact{ClosedPostOpTFourIndGainMean} \\
  \texttt{ctorsearch} & 2 & an explicit constructor term replaced by a closer from a fixed menu of core syntax & \fact{ClosedPostOpCtorsearchSteps} & \fact{ClosedPostOpCtorsearchRows} & \fact{ClosedPostOpCtorsearchGainMean} \\
  \texttt{argon\_funind} & 3 & \texttt{induction} becomes \texttt{fun\_induction}, closers adapted & \fact{ClosedPostOpArgonfunindSteps} & \fact{ClosedPostOpArgonfunindRows} & \fact{ClosedPostOpArgonfunindGainMean} \\
  \texttt{argon\_hbelab} & 3 & elaboration-cost rewrites (\texttt{rw} to \texttt{rewrite}, \texttt{simp only} to \texttt{dsimp only}, ...) & \fact{ClosedPostOpArgonhbelabSteps} & \fact{ClosedPostOpArgonhbelabRows} & \fact{ClosedPostOpArgonhbelabGainMean} \\
  \texttt{argon\_hpcert} & 3 & Lean's own \texttt{?} certificates for bare simp and grind calls; list pruning & \fact{ClosedPostOpArgonhpcertSteps} & \fact{ClosedPostOpArgonhpcertRows} & \fact{ClosedPostOpArgonhpcertGainMean} \\
  \texttt{sonnet\_listdrop} & 3 & single deletions (a list item, \texttt{only}, a leaf step), screened, then their union & \fact{ClosedPostOpSonnetlistdropSteps} & \fact{ClosedPostOpSonnetlistdropRows} & \fact{ClosedPostOpSonnetlistdropGainMean} \\
  \texttt{sonnet\_lintdrop} & 3 & simp arguments that Lean's unused-argument linter reports, dropped & \fact{ClosedPostOpSonnetlintdropSteps} & \fact{ClosedPostOpSonnetlintdropRows} & \fact{ClosedPostOpSonnetlintdropGainMean} \\
  \texttt{opus\_hoist} & 3 & a step repeated across goals becomes one \texttt{any\_goals}/\texttt{all\_goals} & \fact{ClosedPostOpOpushoistSteps} & \fact{ClosedPostOpOpushoistRows} & \fact{ClosedPostOpOpushoistGainMean} \\
  \texttt{opus\_stepdrop} & 3 & a preparation step that does not end its block, dropped & \fact{ClosedPostOpOpusstepdropSteps} & \fact{ClosedPostOpOpusstepdropRows} & \fact{ClosedPostOpOpusstepdropGainMean} \\
  \texttt{opus\_rundrop} & 3 & a run of steps before a block's closer, dropped & \fact{ClosedPostOpOpusrundropSteps} & \fact{ClosedPostOpOpusrundropRows} & \fact{ClosedPostOpOpusrundropGainMean} \\
  \texttt{opus\_scaffold} & 3 & a \texttt{have}/\texttt{obtain} whose names are never used again, removed & \fact{ClosedPostOpOpusscaffoldSteps} & \fact{ClosedPostOpOpusscaffoldRows} & \fact{ClosedPostOpOpusscaffoldGainMean} \\
  \texttt{opus\_armfold} & 3 & a split whose arms close alike becomes one \texttt{<;>} combinator & \fact{ClosedPostOpOpusarmfoldSteps} & \fact{ClosedPostOpOpusarmfoldRows} & \fact{ClosedPostOpOpusarmfoldGainMean} \\
  \texttt{opus\_goalmerge} & 3 & parallel goal blocks merged into one combinator & \fact{ClosedPostOpOpusgoalmergeSteps} & \fact{ClosedPostOpOpusgoalmergeRows} & \fact{ClosedPostOpOpusgoalmergeGainMean} \\
  \texttt{opus\_patfuse} & 3 & a destructuring step fused into the pattern that bound its hypothesis & \fact{ClosedPostOpOpuspatfuseSteps} & \fact{ClosedPostOpOpuspatfuseRows} & \fact{ClosedPostOpOpuspatfuseGainMean} \\
  \texttt{sonnet\_surface} & 3 & token-saving surface rewrites (field notation, inferable binder types dropped) & \fact{ClosedPostOpSonnetsurfaceSteps} & \fact{ClosedPostOpSonnetsurfaceRows} & \fact{ClosedPostOpSonnetsurfaceGainMean} \\
  \texttt{sol\_harvest} & 3 & simplifier calls run as \texttt{?} tactics in one compile; their edits proposed & \fact{ClosedPostOpSolharvestSteps} & \fact{ClosedPostOpSolharvestRows} & \fact{ClosedPostOpSolharvestGainMean} \\
  \texttt{v4\_unhyg} & 4 & renames under one \texttt{unhygienic}, with the names Lean prints & \fact{ClosedPostOpVFourUnhygSteps} & \fact{ClosedPostOpVFourUnhygRows} & \fact{ClosedPostOpVFourUnhygGainMean} \\
  \texttt{hb\_hotctl} & 4 & elaboration controls on the costliest steps of a per-step heartbeat profile & \fact{ClosedPostOpHbhotctlSteps} & \fact{ClosedPostOpHbhotctlRows} & \fact{ClosedPostOpHbhotctlGainMean} \\
  \texttt{v4\_patelide} & 4 & a flat destructuring step becomes \texttt{rcases} under \texttt{unhygienic} & \fact{ClosedPostOpVFourPatelideSteps} & \fact{ClosedPostOpVFourPatelideRows} & \fact{ClosedPostOpVFourPatelideGainMean} \\
  \texttt{v4\_rescreen} & 4 & re-runs version~3's screens with copies that keep the declaration's context & \fact{ClosedPostOpVFourRescreenSteps} & \fact{ClosedPostOpVFourRescreenRows} & \fact{ClosedPostOpVFourRescreenGainMean} \\
  \texttt{hb\_hotelab} & 4 & elaboration-cost rewrites on the profiled hot steps & \fact{ClosedPostOpHbhotelabSteps} & \fact{ClosedPostOpHbhotelabRows} & \fact{ClosedPostOpHbhotelabGainMean} \\
}

\newcommand{\ClosedUploadRows}{%
  \fact{ClosedUphSubTwoZeroSentUtc} & run~1's output (a probe) & \fact{ClosedUphSubTwoZeroPrediction} & \fact{ClosedUphSubTwoZeroCombined} & \fact{ClosedUphSubTwoZeroL} & \fact{ClosedUphSubTwoZeroH} \\
  \fact{ClosedUphSubTwoTwoSentUtc} & run~1's state early in the run, uploaded on purpose as a lower placeholder score (a probe) & \fact{ClosedUphSubTwoTwoPrediction} & \fact{ClosedUphSubTwoTwoCombined} & \fact{ClosedUphSubTwoTwoL} & \fact{ClosedUphSubTwoTwoH} \\
  \fact{ClosedUphSubTwoThreeSentUtc} & run~2's state while it ran (a probe) & \fact{ClosedUphSubTwoThreePrediction} & \fact{ClosedUphSubTwoThreeCombined} & \fact{ClosedUphSubTwoThreeL} & \fact{ClosedUphSubTwoThreeH} \\
  \fact{ClosedUphSubTwoFiveSentUtc} & run~3's state while it ran (a probe) & \fact{ClosedUphSubTwoFivePrediction} & \fact{ClosedUphSubTwoFiveCombined} & \fact{ClosedUphSubTwoFiveL} & \fact{ClosedUphSubTwoFiveH} \\
  \fact{ClosedUphSubTwoSevenSentUtc} & run~2's output & \fact{ClosedUphSubTwoSevenPrediction} & \fact{ClosedUphSubTwoSevenCombined} & \fact{ClosedUphSubTwoSevenL} & \fact{ClosedUphSubTwoSevenH} \\
  \fact{ClosedUphSubThreeZeroSentUtc} & run~1's output after a first post-step with the kernel-check skip on some problems & \fact{ClosedUphSubThreeZeroPrediction} & \fact{ClosedUphSubThreeZeroCombined} & -- & -- \\
  \fact{ClosedUphSubThreeTwoSentUtc} & run~3's output after post-step version~2, with the kernel-check skip on some problems & \fact{ClosedUphSubThreeTwoPrediction} & \fact{ClosedUphSubThreeTwoCombined} & \fact{ClosedUphSubThreeTwoL} & \fact{ClosedUphSubThreeTwoH} \\
  \fact{ClosedUphSubThreeFourSentUtc} & run~3's output after post-step version~2, kernel-check skip withdrawn & \fact{ClosedUphSubThreeFourPrediction} & \fact{ClosedUphSubThreeFourCombined} & \fact{ClosedUphSubThreeFourL} & \fact{ClosedUphSubThreeFourH} \\
  \fact{ClosedUphSubFourOneSentUtc} & run~3's output after two passes of post-step version~4, no kernel-check skip: \textbf{the final upload} & \fact{ClosedUphSubFourOnePrediction} & \fact{ClosedUphSubFourOneCombined} & \fact{ClosedUphSubFourOneL} & \fact{ClosedUphSubFourOneH} \\
}

\usepackage{array}                               %
\newcommand{\cf}[1]{\fact{Closed#1}}            %
\newcommand{\stg}[2]{\fact{ClosedRun#1Stage#2Delta}}  %
\newcommand{\usd}[1]{US\$\fact{Closed#1}}
\newcommand{\ci}[2]{[\fact{Closed#1},\,+\fact{Closed#2}]}   %
\newcommand{\cih}[2]{[+\fact{Closed#1},\,+\fact{Closed#2}]} %

\title{Compile-Gated Lean Proof Refactoring with Enterprise Agent Practices: A Model per Step under a US\$\vxcore{RuleClosedCapUsd} Cap, Followed by Programmatic Rewrite Rules}

\author{%
  Bobber Cheng \\
  Vexorium \\
  \texttt{bobber.cheng@gmail.com} \\
}

\newcommand{\exampleref}{Appendix~\ref{sec:example}}%
\begin{document}

\maketitle

\begin{abstract}
Our closed-track entry (Arena username \texttt{Vexorium-Agent}) scored \cf{UploadBoardCombined} points on the final board
($L=\cf{UploadBoardL}$, $H=\cf{UploadBoardH}$, $Z=\cf{UploadBoardZ}$) over the \vxcore{BenchRows} final problems, against
\vxcore{SeedScore} for the reference proofs, \cf{FinalRank}th at the deadline (top: \cf{FinalBoardTopScore}), for
\usd{RunThreeSpendListTotalUsd} at list prices with every call counted (per problem: mean \usd{RunThreeSpendListMeanUsd}, at most
\usd{RunThreeSpendListMaxUsd}; US\$\vxcore{RuleClosedCapUsd} cap). Three practices carry the entry. First, each paid stage runs the model that
measured best per dollar for it among the \cf{DevModelsMeasured} we tested in our own harness: GPT-6.1 Sol the early sessions, GPT-6.1
Terra the later ones and the tail at \cf{RunThreeCostPerTurnTerraOverSolPct}\,\% of Sol's cost per turn, and Gemini~4 Argon the review and repair; Sol also runs the style rewrite. Second, what our development and
research runs learned ships as compiler-checked programmatic rewrite rules rather than prompt text
($+\cf{RunThreeZeroCostPrereleaseDelta}$ points from rules fixed before the release; $+\cf{RunThreeZeroCostPostreleaseDelta}$ and
$+\cf{PostGain}$, in sample, from passes built or switched on after the release). Third, every call is recorded and reviewed before
each change (all \cf{ClosedFinalUploads} board scores came within
\cf{ArenaMinusMeterMin} to $+\cf{ArenaMinusMeterMax}$ points of our meter's prediction). The harness, gate and many of these rules also serve our open-track entry. We
report the full bill and the spread of three runs.
\end{abstract}

\section{Introduction}
\label{sec:intro}
\providecommand{\exampleref}{Section~\ref{sec:example}}%
The Lean Refactor Arena~\citep{lu2026lean} asks for better proofs of theorems that already have one. Each of its
\vxcore{BenchRows} final problems is a theorem of a Lean~4 library~\citep{moura2021lean4} with a working reference proof; an entry
returns a proof per problem with the statement unchanged, and the Arena rewards fewer tokens, fewer heartbeats (Lean's count of
elaboration work) and compiling unchanged on every toolchain the problem lists (Section~\ref{sec:background}). We entered both
tracks, open weights and hosted models, with one harness, built on a design principle for each technical problem:
\begin{enumerate}\setlength{\itemsep}{1pt}\setlength{\parskip}{0pt}
\item \textbf{A two-objective score on a noisy meter.} Saving a token is worth as much as saving $B/T$ heartbeats ($T$, $B$: the
  reference proof's tokens and heartbeats), and $B/T$ runs from \vxcore{BenchHbPerTokenMin} to \vxcore{BenchHbPerTokenMax} across
  the final problems (median \vxcore{BenchHbPerTokenMedian}); heartbeat readings vary between runs. \emph{Measure, do not guess:}
  every candidate is read on a local copy of the Arena's meter, back to back with the proof it would replace, and kept only by a
  margin above the meter's spread (Section~\ref{sec:gate}).
\item \textbf{Every listed toolchain.} \vxcore{BenchMultipinRows} of the \vxcore{BenchRows} problems list several toolchains (up
  to \vxcore{BenchMaxPins}, \vxcore{BenchToolchainOldest} to \vxcore{BenchToolchainNewest}), and a proof that fails on one loses
  points. \emph{Keep nothing whose $Z$ has not been measured on all of them.}
\item \textbf{A hard budget}: US\$\vxcore{RuleClosedCapUsd} per problem at list prices, or
  \vxcore{RuleOpenGpus}\,$\times$\,A100 \vxcore{RuleOpenGpuGb}\,GB for \vxcore{RuleOpenHours}\,h. \emph{Use the model only where
  rules cannot do the work:} mechanical rewrites run as model-free CPU stages behind the same gate (Section~\ref{sec:endstage}).
\item \textbf{Few public rows}: \vxcore{WarmupRows} warm-up problems before a final benchmark with \vxcore{BenchNewRows} new problems
  and \vxcore{BenchNewToolchains} new toolchains. \emph{General mechanisms only:} no stage is keyed to a problem, every final proof
  starts from the released reference proof, and work after the release is disclosed.
\end{enumerate}
Both reports share Section~\ref{sec:background}, Sections~\ref{sec:gate} and~\ref{sec:endstage}, \exampleref{} and
part of Section~\ref{sec:lessons}; each entry's agent stage, models, budget and results are its own.

\paragraph{This report: the closed track.} It covers our hosted-model entry; the open-weights entry has its own report. Our thesis:
\emph{under a US\$\vxcore{RuleClosedCapUsd} cap per problem, budget steps each run by the model that measured best per dollar in our own
loop, rewrite rules learned from our own development and research runs and shipped as compiler-checked programmatic stages, and every call
recorded and reviewed take the \vxcore{BenchRows} final proofs from \vxcore{SeedScore} to \cf{UploadBoardCombined} at a mean
\usd{RunThreeSpendListMeanUsd} per problem.} ``Programmatic'' (model-free below) means that the rules call no model and sample nothing;
their reruns are still not byte-identical (Section~\ref{sec:endstage}). Contributions: a model per budget step measured against its
alternatives (Section~\ref{sec:decisions}), a cap enforced in code with a bill of all \cf{RunThreePaidCalls} calls of the uploaded run
(Sections~\ref{sec:agent} and~\ref{sec:budget}, Appendix~\ref{app:perproblem}), and gated rules learned from our own runs, with their
provenance and what they gave our open entry (Sections~\ref{sec:endstage} and~\ref{sec:lessons}, Appendices~\ref{app:devlessons}
and~\ref{app:ops}). The design principles in Section~\ref{sec:lessons} come from enterprise agent systems from Vexorium.
Our code is public at \url{https://github.com/bobbercheng/lean-refactor-closed} (Appendix~\ref{app:pins}).

\section{Background}
\label{sec:background}
\paragraph{Benchmark.} The final benchmark~\citep{lu2026lean} has \vxcore{BenchRows} problems, \vxcore{BenchRowsPerLibrary} from
each of \vxcore{BenchLibraries} sources: theorems of ArkLib~\citep{arklib}, CSLib~\citep{cslib}, LeanPool~\citep{leanpool},
PhysLib~\citep{physlib} and Strata~\citep{strata}, compiled inside their repositories, and
PutnamBench~\citep{tsoukalas2024putnambench} problems, compiled against Mathlib~\citep{mathlib2020}. A problem gives the statement,
the reference proof and its token count (\vxcore{BenchRefTokensMin} to \vxcore{BenchRefTokensMax}, median
\vxcore{BenchRefTokensMedian}), and its toolchains, each pinned to a commit of the source repository. It was released on
\vxcore{ReleaseUtc}; \vxcore{WarmupSharedRows} of its problems are among the \vxcore{WarmupRows} public warm-up problems.

\paragraph{Score.} Each problem scores the mean of its length reduction $L$ (proof-body tokens against the reference proof, counted
by the Arena's lexer), heartbeat reduction $H$ (Lean's \texttt{\#count\_heartbeats} against the reference proof) and zero-shot
compatibility $Z$ (the share of the listed toolchains on which the proof compiles unchanged), all in percent~\citep{lu2026lean};
neither reduction is floored. A problem that is not submitted, fails to compile, changes the statement or uses a forbidden
construct scores zero, and the board score is the mean over all problems. The reference proof itself scores \vxcore{SeedScore}
($L=\vxcore{SeedL}$, $H=\vxcore{SeedH}$, $Z=\vxcore{FullZ}$). On every final problem the first-listed toolchain is the newest; our
meters read heartbeats there.

\paragraph{Forbidden constructs.} The Arena rejects a proof whose text, comments stripped, matches a fixed list: commands that
run code (\texttt{\#eval}, \texttt{run\_tac}, \dots), escape hatches (\texttt{sorry}, \texttt{axiom}, \texttt{native\_decide},
\texttt{unsafe}, \texttt{IO}, \dots), new syntax, simprocs, auxiliary declarations, \texttt{attribute} commands, and identifiers
that mention heartbeats (except \texttt{maxHeartbeats}). Both entries check a copy of the full list before compiling.

\paragraph{Rules that shaped the design.} Closed-track entries call every model through OpenRouter and spend at most
US\$\vxcore{RuleClosedCapUsd} per problem at the provider's list prices; open-track entries use public weights on at most
\vxcore{RuleOpenGpus}\,$\times$\,A100 \vxcore{RuleOpenGpuGb}\,GB and finish the benchmark run within \vxcore{RuleOpenHours}\,h.
Entries upload proofs and code with a README that regenerates them, and the organizers rerun top entries. In answers on the
competition's Discord, the organizers added that final proofs must be refactored from the released reference proofs, not from
proofs or hints found earlier. Both our pipelines start every problem from its reference proof; what each code package carries over
from development is in its report's Disclosures.

\paragraph{What binds the closed track.} This entry runs under the closed-track rules: models only through OpenRouter, at most
US\$\vxcore{RuleClosedCapUsd} per problem at list prices, every call counted, retries and discarded candidates included. Its budget is
dollars, not hours: a session turn cost about US\$\cf{RunThreeCostPerTurnTerraLadderUsd} (Terra) to US\$\cf{RunThreeCostPerTurnSolLadderUsd}
(Sol) in the uploaded run while CPU work costs nothing, so every call is priced at its worst case before it is sent and work that needs no
model goes to \$0 CPU stages.

\section{Approach}
\label{sec:approach}
Figure~\ref{fig:pipeline} shows the pipeline that follows from those limits: budget steps of sessions, each step run by the model
measured best for it (Section~\ref{sec:agent}), a gate that decides what any stage may keep (Section~\ref{sec:gate}, with each stage's bar),
and model-free CPU stages that make the mechanical rewrites at \$0 (Section~\ref{sec:endstage}). A run is one command of the code
archive; the post-step is two more on its output. Parts built or switched on after the release are marked in the text and listed in
Appendix~\ref{app:disclosures}.

\begin{figure}[t]
  \centering
  \begin{tikzpicture}[font=\scriptsize, x=1cm, y=1cm,
    stage/.style={draw=black!60, rounded corners=1.5pt, align=flush center, minimum height=1.42cm, inner sep=2pt, anchor=north west},
    sub/.style={draw=vxBlue!55, fill=white, rounded corners=1pt, align=flush left, inner sep=2pt, anchor=north west},
    cell/.style={draw=vxBlue!35, fill=vxBlue!5, rounded corners=1pt, align=flush center, inner sep=1pt, anchor=north west, minimum height=0.87cm},
    lane/.style={align=flush left, inner sep=0pt, anchor=north west, text width=1.45cm},
    sol/.style={fill=vxBlue!18}, terra/.style={fill=vxGreen!30}, argon/.style={fill=vxOrange!25},
    cln/.style={draw=vxGray, fill=white, circle, inner sep=0.6pt, font=\tiny},
    arr/.style={-{Stealth[length=4pt]}, black!70, line width=0.6pt},
    sar/.style={-{Stealth[length=3pt]}, black!60, line width=0.5pt}]
  \fill[black!4, draw=black!60, rounded corners=1.5pt] (1.89,0) -- (3.98,0) -- (3.98,-1.56) -- (13.9,-1.56) -- (13.9,-5.70) -- (0,-5.70)
    -- (0,-1.56) -- (1.89,-1.56) -- cycle;
  \node[stage, fill=black!6, text width=1.55cm] (rows) at (0,0) {\textbf{Problems}\\ \vxcore{BenchRows} reference proofs, pinned toolchains};
  \node[stage, draw=none, text width=1.95cm] (agent) at (1.89,0) {\textbf{Budget steps}\\ resumed sessions to \$\cf{RunThreeLadderTopUsd} a problem};
  \node[stage, argon, text width=2.2cm] (rev) at (4.18,0) {\textbf{Review} (\$0 stack)\\ Gemini~4 Argon:\\ \emph{review, repair};\\ improved \cf{RunThreeFigReviewImprovedRows} problems};
  \node[stage, terra, text width=1.85cm] (tail) at (6.72,0) {\textbf{Tail}\\ GPT-6.1 Terra: \emph{tail sessions}; \cf{RunThreeSessionsTail} on \cf{RunThreeFigTailRows} problems, \cf{RunThreeTailAccepted} kept};
  \node[stage, fill=white, text width=1.85cm] (end) at (8.91,0) {\textbf{End stage} (CPU),\\ {\setlength{\fboxsep}{0.8pt}\colorbox{vxBlue!18}{Sol \emph{style rewrite}}}\kern-0.8pt,\\ end passes};
  \node[stage, fill=white, text width=1.35cm] (post) at (11.1,0) {\textbf{Post-step}\\ (CPU) \cf{PostOpsRan} ops, two passes};
  \node[stage, fill=black!6, text width=0.95cm] (up) at (12.79,0) {\textbf{Upload}\\ board\\ \cf{UploadBoardCombined}};
  \foreach \a/\b in {rows/agent, agent/rev, rev/tail, tail/end, end/post, post/up} \draw[arr] (\a.east |- 0,-0.71) -- (\b.west |- 0,-0.71);
  \draw[decorate, decoration={brace, amplitude=3pt}, black!60] (agent.north west |- 0,0.05) --
    node[above=2pt] {run~3: one command; \cf{RunThreeShards} shards on one \cf{RunThreeHostVcpu}-vCPU host, \cf{RunThreeWallH}\,h; output mean \cf{RunThreeStoredMean} on our meter} (end.north east |- 0,0.05);
  \node[lane] at (0.09,-1.68) {\textbf{Budget steps}\\ spend picks the model};
  \node[sub, sol, text width=2.3cm, minimum height=1.28cm] (s1) at (1.62,-1.66) {\textbf{Step 1}, to \$\cf{RunThreeRungOneUsd}\\ Sol: \emph{early sessions}\\ \cf{RunThreeSessionsSol} sessions, \cf{RunThreeTurnsPerSessionSol} turns\\ ends at a \textbf{checkpoint}};
  \node[sub, terra, text width=2.3cm, minimum height=1.28cm] (s2) at (4.4,-1.66) {\textbf{Step 2}, to \$\cf{RunThreeRungTwoUsd}\\ Terra: \emph{later sessions}\\ \cf{RunThreeSessionsTerraRungOneNineZero} sessions, \cf{RunThreeFigTurnsPerSessionRungOneNineZero} turns\\ ends at a \textbf{checkpoint}};
  \node[sub, terra, text width=2.3cm, minimum height=1.28cm] (s3) at (7.18,-1.66) {\textbf{Step 3}, to \$\cf{RunThreeLadderTopUsd}\\ Terra: \emph{later sessions}\\ \cf{RunThreeSessionsTerraRungTwoSevenFive} sessions, \cf{RunThreeFigTurnsPerSessionRungTwoSevenFive} turns\\ then cleanup, review};
  \node[sub, dashed, text width=3.76cm, minimum height=1.28cm] (sch) at (9.92,-1.66) {\textbf{Scheduler}, each step: \cf{RunThreeShardWorkers} sessions at once a shard, one per problem; a first session waits for its auto-brief, then fewest sessions on the step, least spend};
  \draw[sar] (s1.east) -- node[cln] {c} (s2.west);
  \draw[sar] (s2.east) -- node[cln] {c} (s3.west);
  \node[lane] at (0.09,-3.14) {\textbf{One session}\\ steps and tail: \cf{RunThreeSessionsTotal}, mean \cf{RunThreeFigSessionMeanMinutes}\,min};
  \node[sub, text width=1.95cm, minimum height=1.3cm] (open) at (1.62,-3.12) {\textbf{Open} from state; auto-brief start on \cf{RunThreeAutobriefStartRows} problems; library leads, cards};
  \node[sub, minimum width=7.9cm, minimum height=1.3cm] (loop) at (3.84,-3.12) {};
  \node[anchor=north west, inner sep=0pt] at (3.91,-3.16) {\textbf{Tool loop}, at most \cf{RunThreeSessionTurnBudget} turns (\cf{RunThreeTurnsTotal} in all): model turn, then tool calls};
  \node[cell, text width=1.6cm] (c1) at (3.91,-3.46) {\textbf{Compile}, meter, \textbf{batches} of 1--\cf{RunThreeBatchCandsMax}\\ \cf{RunThreeFigCompileCalls} calls};
  \node[cell, text width=1.45cm] (c2) at (5.64,-3.46) {\textbf{Scratch Lean}: goals, profiles\\ \cf{RunThreeFigScratchCalls} calls};
  \node[cell, text width=1.25cm] (c3) at (7.22,-3.46) {\textbf{Library} search\\ \cf{RunThreeFigLibraryCalls} calls};
  \node[cell, text width=1.6cm] (c4) at (8.6,-3.46) {\textbf{Bank} a higher score\\ \cf{RunThreeFigBanks} banks};
  \node[cell, text width=1.2cm] (c5) at (10.33,-3.46) {\textbf{Notes} file\\ \cf{RunThreeFigNotesSessions} sessions};
  \node[sub, text width=1.82cm, minimum height=1.3cm] (stop) at (11.86,-3.12) {\textbf{Stop}: finish \cf{RunThreeStopFinish}\\ turn budget \cf{RunThreeStopTurnBudget}\\ budget gate \cf{RunThreeStopBudgetGate}};
  \draw[sar] (open.east) -- (loop.west);
  \draw[sar] (loop.east) -- (stop.west);
  \node[lane] at (0.09,-4.56) {\textbf{Between sessions}\\ (CPU, \$0)};
  \node[draw=vxBlue!55, fill=black!10, rounded corners=1pt, anchor=north west, minimum width=12.2cm, minimum height=0.32cm, inner sep=1pt, align=center] (state) at (1.62,-4.54)
    {\textbf{Problem state} on disk: best verified proof, compiled candidates, history, notes};
  \node[sub, text width=5.9cm, minimum height=0.56cm] (cl) at (1.62,-4.98) {\tikz[baseline=-0.6ex]\node[draw=vxGray, fill=white, circle, inner sep=0.6pt, font=\tiny, minimum height=0pt, minimum width=0pt]{c}; \textbf{Cleanup} after each step (model-free passes, rules): better proof on \cf{RunThreeFigCleanZeroAdopted}, \cf{RunThreeFigCleanOneAdopted} and \cf{RunThreeFigCleanTwoAdopted} problems};
  \node[sub, text width=5.9cm, minimum height=0.56cm] (wb) at (7.78,-4.98) {\textbf{Write-back} into the sessions' state: \cf{RunThreeFigWritebackCleanZero}\,+\,\cf{RunThreeFigWritebackCleanOne} checkpoint gains, \cf{RunThreeFigWritebackTailStart}~bests at the tail start (\cf{RunThreeWritebackRows} problems)};
  \foreach \n in {c4, c5} \draw[sar] (\n.south) -- (\n.south |- state.north);
  \draw[sar] (state.north -| open.south) -- (open.south);
  \draw[sar] (wb.north) -- (wb.north |- state.south);
  \draw[sar] (cl.east) -- (wb.west);
  \node[draw=black!50, fill=black!3, rounded corners=1.5pt, anchor=north west, text width=13.68cm, align=flush left, inner sep=3pt] (gate) at (0,-5.80)
    {\textbf{Cap gate} on every paid call: its worst case (all input at the cache-write rate, the full output cap) is priced before sending
    and must fit the session's budget or, for other calls, the room under US\$\cf{CapUsdPerProblem}; a turn's output cap shrinks to fit, and a session ends when the smallest cap no longer does. \textbf{Accept
    gate} (Section~\ref{sec:gate}) on every output: exact statement; forbidden patterns; no \texttt{sorry}; paired rule (no lower $Z$) or score bar.\\[1pt]
    \mbox{\tikz\node[fill=vxBlue!18,draw=vxBlue!55,minimum size=0.18cm,inner sep=0pt]{}; GPT-6.1 Sol: early sessions, style rewrite}\enspace
    \mbox{\tikz\node[fill=vxGreen!30,draw=vxBlue!55,minimum size=0.18cm,inner sep=0pt]{}; GPT-6.1 Terra: later sessions, tail}\enspace
    \mbox{\tikz\node[fill=vxOrange!25,draw=black!60,minimum size=0.18cm,inner sep=0pt]{}; Gemini~4 Argon: review, repair}\enspace
    \mbox{\tikz\node[cln]{c}; cleanup}};
  \end{tikzpicture}
  \caption{The closed pipeline (run~3 and the upload), budget steps opened up. Counts: run~3's totals over the
    \vxcore{BenchRows} problems; turns: means per session; budget steps: cumulative spend per problem at list prices. A checkpoint
    ends steps 1 and 2 once a shard's problems all reach the step's top: cleanup, write-back, go on or stop
    (stopped: \cf{RunThreeFigCheckpointZeroStopRows} after step 1, no headroom; \cf{RunThreeFigCheckpointOneStopRows} after step 2). Not
    drawn: a \$0 crossover before the review (adopted on \cf{RunThreeFigCrossAdopted}). Post-step ops: Appendix~\ref{app:ops} (no
    kernel-check skip).}
  \label{fig:pipeline}
\end{figure}

\subsection{The budget-laddered agent stage}
\label{sec:agent}
\paragraph{Budget steps of resumed sessions.} Each problem keeps its state on disk; a session, a tool-using conversation, opens from it
and runs the tool loop of Figure~\ref{fig:pipeline}. Unlike the open package, closed sessions also have a library-search tool, and every
opening carries library leads and technique cards, most also pattern cards (\cf{RunThreeCardsPatternSessions} of
\cf{RunThreeSessionsTotal}; openings: median \cf{RunThreeOpeningWordsMedian} words). Batched checks (one call compiles up to
\cf{RunThreeBatchCandsMax} full candidates in turn and banks any that beats the best) carried \cf{RunThreeBatchCandSharePct}\,\% of
the checked candidates (\cf{RunThreeBatchCalls} calls, \cf{RunThreeBatchCandsMean} candidates each), yet turns with one made only
\cf{RunThreeBatchNewbestTurnPct}\,\% of the new bests (\cf{RunThreeBatchGainSharePct}\,\% of their summed gain): these models found most
gains one candidate at a time. Every problem's first session starts from its reference proof, shown with the CPU auto-brief's better start
if found (\cf{OpenGapHeadStartRows} problems start above their reference, $+\cf{OpenGapHeadStart}$ on the mean before any model
edit); later sessions resume from the stored best, into which checkpoint cleanups and, before the tail, the current best are
written back (\cf{RunThreeWritebackCount} times on \cf{RunThreeWritebackRows} problems); none restarts from the reference.

\paragraph{Changes for these models.} Settings were measured in our harness, not carried over. At extra-high effort Sol fit
\cf{OpenSolXhighTurns} turns per problem in the first budget step against \cf{OpenSolMediumTurns} at medium and scored
$+\cf{DlXhighLatestGain}$ \ci{DlXhighLatestLo}{DlXhighLatestHi} against the latest \cf{DlXhighLatestN} medium draws (a development
note): no measured gain, so it runs at medium. Terra runs the later steps at \cf{RunThreeCostPerTurnTerraOverSolPct}\,\% of Sol's cost per turn,
and Argon cost more as the main model for the same score (Table~\ref{tab:decisions}). These models wrote \cf{OpenGlmClosedOutputPerTurn} output tokens per turn (GLM-5.3-Flash:
\cf{OpenOutputTokensPerTurn}), so unlike the open entry they needed no budget forcing or context cap.

\paragraph{Review, tail, end.} Argon then reviews with the room left (the cap minus the problem's spend):
\cf{RunThreeReviewRounds} rounds of \cf{RunThreeReviewK} proposed edits; a failing edit gets one repair turn (at most
\cf{RunThreeReviewRepairsPerRound} per round), and a round's compiling edits are also tried together (the stack, a \$0 compile). Room still left
buys up to \cf{RunThreeTailMaxSessions} Terra sessions (the tail). The end stage (Section~\ref{sec:endstage}) also polishes the best
proof of one other approach among the problem's candidates (the approach pick, at most \cf{RunThreeApproachSeconds}\,s). The style
rewrite asks Sol, with at most \usd{RunThreeStylerwBudgetUsd} and \cf{RunThreeStylerwWallS}\,s, for the proof in the short style the
score rewards, guided by \cf{DiscExemplarPairs} fixed before/after example pairs, and keeps it only on a back-to-back gain of \cf{OpenGateOldBar}.

\paragraph{The cap in code.} The cap gate's worst case (Figure~\ref{fig:pipeline}) is the whole input at the highest input rate
(cache write) plus the output cap (Sol \cf{ModelSolGatePrices}, Terra \cf{ModelTerraGatePrices}, Argon \cf{ModelArgonGatePrices}
US\$/M input / output / cache-read / cache-write tokens; Argon at its announced later price, Section~\ref{sec:budget}). A call is metered at the higher of OpenRouter's bill and the list
price; a possibly billed failed attempt books its worst case; spend survives restarts; every request carries a
\texttt{max\_price} pin.

\paragraph{Observability and review.} Every model call is recorded with its tokens, billed and metered cost and, for session turns,
the serving provider; every session with the problem's spend and score before and after; every stage's proof with its paired reading.
During each run a live trace (W\&B) plotted the mean score, spend, tokens and API errors. Tables~\ref{tab:budget}
and~\ref{tab:ablation} and Appendix~\ref{app:perproblem} are sums of these records, and every upload's board score is checked against our
meter's prediction (Section~\ref{sec:gate}). The author reviewed the records after each test and run and decided each row of
Table~\ref{tab:decisions} on them; the records also exposed defects that tests had missed, such as a first session that did not see the
auto-brief's start (fixed in the frozen code, Disclosures), and Appendix~\ref{app:devlessons} lists more.

\subsection{The verification gate}
\label{sec:gate}
Every stage of both pipelines replaces a proof under one acceptance rule or its own score bar (below);
the rule, the forbidden-pattern filter and the token counter are the same code in both archives. Under the rule a candidate must
(i)~start with the exact statement followed by \texttt{:=};
(ii)~pass our copy of the forbidden-pattern filter;
(iii)~compile on no fewer listed toolchains than the current proof (on more, a higher score suffices; the post-step: on all);
(iv)~not depend on \texttt{sorry}: every compile forces Lean's \texttt{warn.sorry} on (PhysLib's lakefile turns it off, which would
accept a proof reaching \texttt{sorryAx} without the word, e.g.\ through \texttt{stop}), and the post-step also audits
every proof it adopts with \texttt{\#print axioms}, except in files that are Lean \texttt{module}s, where only the \texttt{sorry}
check runs;
(v)~win a paired reading: measured back to back with the current proof on the same meter and compared unrounded, its worse
heartbeat reading against the current proof's better one, it must save a token or at least \vxcore{AcceptHbEps} heartbeats (above
the meter's repeat spread) and either dominate (no more tokens, no more heartbeats, a higher score) or gain at least
\vxcore{AcceptMinGain} points. Errors and timeouts keep the current proof.

\paragraph{The meter.} Tokens are counted by a re-implementation of the lexer of ProofOptimizer~\citep{gu2025proofoptimizer} (the
Arena's worker is not public); it reproduces the released counts of all \vxcore{TokenizerWarmupMatch} warm-up reference proofs and
of \vxcore{TokenizerFinalMatch} of the \vxcore{BenchRows} final ones, and is at most \vxcore{TokenizerFinalMaxDiff} tokens off on
the rest. Heartbeats are Lean's counter read around the declaration's elaboration with \texttt{Elab.async} off, on the first-listed
toolchain (Mathlib's \texttt{\#count\_heartbeats} reads the same counter, far more slowly on some declarations). The release had no
reference heartbeats, so both entries measured them with this meter plus a constant offset fitted on the warm-up problems: within
\vxcore{RefhbWarmupMaxErrPct}\,\% of the \vxcore{RefhbWarmupRows} published warm-up references (mean
\vxcore{RefhbWarmupMeanErrPct}\,\%), and the two entries' files, measured on different machines, differ by at most
\vxcore{RefhbTracksMaxDiffPct}\,\% on any final problem. Over this entry's \vxcore{MeterUploads} scored uploads with a recorded
prediction, the board score minus our prediction ranged from \vxcore{MeterDiffMin} to \vxcore{MeterDiffMax} points
(\vxcore{MeterDiffUpload} for the final upload).

\paragraph{Where the bar sits.} A strict bar drops good ideas: in our earlier hosted-model pipeline a fixed step of \cf{OpenGateOldBar}
points discarded gains below it (\cf{OpenGateOldBarSumLo}--\cf{OpenGateOldBarSumHi} points in all) on
\cf{OpenGateOldBarRowsLo}--\cf{OpenGateOldBarRowsHi} held-out problems. So the margins follow the meter's spread: the end stage, end passes and post-step keep a proof
on the paired rule above; the tail and the style rewrite read old and new back to back against \cf{OpenGateOldBar} as a signal bar; the
\$0 cleanups between budget steps and the review re-measure on one meter against that bar; and a session banks a higher
score, $Z$ included, from one reading. Our open entry inherited these
margins.

\subsection{The model-free end stage and post-step}
\label{sec:endstage}
Both entries rewrite proofs without a model in two places: an end stage inside the run and a post-step on the run's export. Neither
calls a model or uses an accelerator, and every candidate goes through the gate. The design is shared and the code largely so; each
entry's variants follow this subsection.

\paragraph{End stage.} It takes the best of the problem's own verified proofs (the reference, each stage's best, the run's pool of
compiled candidates), re-read back to back on the same meter, and searches mechanical rewrites of it: a sweep of step deletions and
merges, crossovers between the problem's proofs at shared tactic boundaries, generic rewrite tools, and a dialect pass that puts the
proof under a single \texttt{unhygienic} so that later steps use the names Lean gives to hypotheses (read from
\texttt{trace\_state}, never guessed). Candidates are screened many to a Lean file and ranked by the problem's own $B/T$.

\paragraph{Post-step.} The post-step (\texttt{harness/open\_post.py}) applies ops, general rewrite rules that read only the proof,
the statement and the problem's own Lean environment. For example, a destructuring pattern becomes projection paths
(\texttt{t.2.1} is a single token to the lexer), an anonymous constructor a leading-dot constructor, structural induction functional
induction, a closer the explicit \texttt{simp only} or \texttt{grind only} call that Lean suggests for its goal, and the costliest
steps of a per-step heartbeat profile get cheaper elaboration settings. By default a problem gets at most \vxcore{PostRounds}
rounds over the ops within \vxcore{PostRowCompiles} compiles and \vxcore{PostRowSeconds}\,s; problems run in parallel, and passes
chain, each on the previous output. A post-step candidate must also raise the problem's rounded score above its stored one, and a
budget stop keeps the last proof that passed the whole gate. No model or sampling is involved, yet reruns are not byte-identical:
ops size their probes from measured compile times (Disclosures).

The post-step is not independent of the final benchmark. Its ops were developed on the final benchmark after its release
(\vxcore{ReleaseUtc}), from research sessions of closed models, a static study of our runs' proofs, and research agents that
compared our open-track and closed-track proofs of the same problems; each op was measured on our own outputs for those problems
(in sample). The ops hold no problem names and no proof text of any problem; Appendix~\ref{app:disclosures} gives the full list in
the words of the code README.

\paragraph{The closed entry's variants.} Its models know more Lean than the open entry's (\cf{OpenLoopClosedFailLo}--\cf{OpenLoopClosedFailHi}\,\%
of their candidates failed to compile, against \cf{OpenLoopCompileFailPctFlashThree}\,\% for GLM-5.3-Flash), yet this entry also leans on
these stages, for two reasons: under the cap a paid turn costs money while a rule costs nothing, and knowledge given as prose was rarely
acted on (Section~\ref{sec:negative}), while a rule that reads names and lemmas from Lean applies every time. Here the dialect pass
runs as a separate gated pass after the style rewrite, then a \emph{hint pass} (\texttt{grind} hint lists minimized; a
simp-list constructor literal replaced by an equal named constant Lean finds) and a \emph{closer pass} (a \texttt{decide}-family closer
moved onto the cheaper hypothesis of a library implication an environment query finds, Lean's suggestion in place of a search tactic,
micro-rewrites). The dialect pass's rules predate the release; all three passes were switched on after it (run~2; run~3 added two closer
variants). The two post-step passes (Appendix~\ref{app:ops}) changed \cf{PostRowsChanged} problems, none lower; the upload has no
\texttt{set\_option} line.

\section{Models}
\label{sec:models}
Table~\ref{tab:models} lists every model the uploaded run called, all via OpenRouter (one key), by id as served from
\cf{RunThreeLaunchUtc} to \cf{RunThreeEndUtc}. OpenRouter recorded OpenAI as provider of all \cf{RunThreeTurnsTotal} session turns;
the \cf{RunThreeNonSessionCalls} other calls carry no provider record. No temperature, \texttt{top\_p} or seed is sent
(provider defaults). No embedding, reranking or judge model is used; a configured fallback of an unused feature never
ran. Development tried \cf{DevModelsMeasured} models in the same harness (Appendix~\ref{app:experiments}); Table~\ref{tab:decisions}
gives the measured trade-offs behind each step's choice.

\begin{table}[t]
  \caption{Models; reasoning effort \cf{RunThreeAllCallsEffort} on every call. Output cap: tokens per call (sessions: range over Sol and Terra; other calls: largest sent), shrunk to fit
    the room under the step's top (sessions) or under the cap (other calls). List price: US\$/M input / output tokens.}
  \label{tab:models}
  \centering\footnotesize
  \begin{tabular}{@{}lllr@{}}
    \toprule
    Role & OpenRouter id & Output cap & List price \\
    \midrule
    Sessions to \$\cf{RunThreeRungOneUsd}; style rewrite & \texttt{\cf{ModelSolOpenrouterId}} &
      \cf{RunThreeTurnMaxTokensMin}--\cf{RunThreeTurnMaxTokensMax}; \cf{RunThreeStylerwMaxTokensMax} & \cf{ModelSolListInUsdPerM} / \cf{ModelSolListOutUsdPerM} \\
    Sessions to \$\cf{RunThreeLadderTopUsd}; tail & \texttt{\cf{ModelTerraOpenrouterId}} &
      \cf{RunThreeTurnMaxTokensMin}--\cf{RunThreeTurnMaxTokensMax} & \cf{ModelTerraListInUsdPerM} / \cf{ModelTerraListOutUsdPerM} \\
    Review; repair & \texttt{\cf{ModelArgonOpenrouterId}} &
      \cf{RunThreeArgonReviewMaxTokensMax}; \cf{RunThreeArgonRepairMaxTokensMax} & \cf{ModelArgonListInUsdPerM} / \cf{ModelArgonListOutUsdPerM} \\
    \bottomrule
  \end{tabular}
\end{table}

\section{Budget accounting}
\label{sec:budget}
The closed track budgets dollars per problem at list prices (Section~\ref{sec:background}). Table~\ref{tab:budget} gives the uploaded
run's spend; a problem's spend is every call made for it, discarded candidates included
(retried turns, turns without a usage record, failed review and style attempts: \cf{RunThreeTurnsRetried},
\cf{RunThreeTurnsUsageEstimated}, \cf{RunThreeReviewFailedAttempts}). Every call's tokens at list prices (Table~\ref{tab:models}, with the cache-read and
cache-write rates) give exactly its metered cost (\cf{RunThreeCallsMeteredAboveList} calls above, \cf{RunThreeCallsMeteredBelowList} below); OpenRouter billed
\cf{RunThreeBilledOverListRatio} of that on all \cf{RunThreeCallsWithBill} calls. Argon was billed at its introductory list price;
the last row re-prices it at Google's announced later price (twice as high). Appendix~\ref{app:perproblem} gives every problem; CPU
stages cost \$0. Development and research calls (not in the table) were paid from Azure and Google Cloud credits, through our own
OpenRouter account, by card (Claude models on Azure's marketplace) and through a Claude subscription; the uncapped research
sessions behind the hint and closer rules (\cf{DiscResearchRows} final problems) cost about \usd{DevResearchUsd} at list prices
(\usd{DevResearchCardUsd} of it by card) and yielded rules, not proofs.

\begin{table}[t]
  \caption{Spend of the uploaded run (run~3) at list prices, all calls included (one call per session turn), \vxcore{BenchRows}
    problems; components rounded.}
  \label{tab:budget}
  \centering\footnotesize
  \begin{tabular}{@{}llrrr@{}}
    \toprule
    Component & Model & Sessions & Calls & US\$ \\
    \midrule
    Sessions to \$\cf{RunThreeRungOneUsd} & GPT-6.1 Sol & \cf{RunThreeSessionsSol} & \cf{RunThreeTurnsSolLadder} & \cf{RunThreeSpendSolSessionsUsd} \\
    Sessions \$\cf{RunThreeRungOneUsd}--\$\cf{RunThreeLadderTopUsd} & GPT-6.1 Terra & \cf{RunThreeSessionsTerraLadder} & \cf{RunThreeTurnsTerraLadder} & \cf{RunThreeSpendTerraSessionsUsd} \\
    Tail & GPT-6.1 Terra & \cf{RunThreeSessionsTail} & \cf{RunThreeTurnsTerraTail} & \cf{RunThreeSpendTerraTailUsd} \\
    Review & Gemini 4 Argon & & \cf{RunThreeReviewCalls} & \cf{RunThreeSpendArgonReviewUsd} \\
    Repairs & Gemini 4 Argon & & \cf{RunThreeRepairCalls} & \cf{RunThreeSpendArgonRepairUsd} \\
    Style rewrite & GPT-6.1 Sol & & \cf{RunThreeStylerwCalls} & \cf{RunThreeSpendSolStylerwUsd} \\
    \midrule
    Total & & & \cf{RunThreePaidCalls} & \cf{RunThreeSpendListTotalUsd} \\
    Per problem: mean / min / \textbf{max} & & & & \cf{RunThreeSpendListMeanUsd} / \cf{RunThreeSpendListMinUsd} / \textbf{\cf{RunThreeSpendListMaxUsd}} \\
    Argon at the cap price: total / max & & & & \cf{RunThreeSpendCapTotalUsd} / \cf{RunThreeSpendCapMaxUsd} \\
    \bottomrule
  \end{tabular}
\end{table}

\section{Reproduction}
\label{sec:repro}
The organizers rerun top entries (Section~\ref{sec:background}); for that, the code archive's \texttt{README.md} and
\texttt{REPLICATE.md} (Appendix~\ref{app:pins}) give each command with its expected output: environments, reference heartbeats, a \$0 scripted-model dry run, the run (\texttt{run\_final.py run}, end
passes on) and the two post-step commands with our op list (the default list has one more op); the archive holds
the style rewrite's example file. On a \cf{RunThreeHostVcpu}-vCPU, \cf{RunThreeHostMemGb}\,GB host the run
took \cf{RunThreeWallH}\,h and the post-step \cf{PostPassOneSeconds}\,s and
\cf{PostPassTwoSeconds}\,s; a rerun takes about as long. Our runs spent \usd{RunOneSpendListTotalUsd}, \usd{RunTwoSpendListTotalUsd} and \usd{RunThreeSpendListTotalUsd}.

\paragraph{Rerun variation.} Provider-default sampling and wall-clock limits make a rerun write other proofs; the uploaded code
ran once. Our full runs (differing in fixes, end passes and host) span \cf{RunsSpan} on our meter
(Table~\ref{tab:ablation}); before the release, \cf{DevWarmRuns} from-reference runs of an older version had an SD of
\cf{DevWarmRunsSd} on the warm-up mean. The post-step's rerun spread on this op list is unmeasured
(Section~\ref{sec:lessons}).

\section{Results and analysis}
\label{sec:results}
\label{sec:headline}
We now follow the score from the reference proofs through the budget steps to the board. Table~\ref{tab:ablation} and
Figure~\ref{fig:curve} (Appendix~\ref{app:perproblem}) split runs into stages; the upload scored \cf{UploadBoardCombined} (our meter:
\cf{UploadStoredMean}). In every run the first stage (auto-brief, then Sol to \usd{RunThreeRungOneUsd};
mean spend \usd{RunThreeSpendFirstStageMeanUsd}) makes \cf{RunThreeFirstStageSharePct}--\cf{RunOneFirstStageSharePct}\,\% of the
gain over the reference; the Terra steps add \cf{RunTwoTerraRungsDelta}--\cf{RunThreeTerraRungsDelta} points, the review
\stg{One}{Review}--\stg{Two}{Review}, tail and style rewrite almost nothing. The \$0 stages frozen before the release add
\cf{RunOneZeroCostPrereleaseDelta}--\cf{RunTwoZeroCostPrereleaseDelta}; the end passes switched on after it add
\cf{RunThreeZeroCostPostreleaseDelta} (run~3) and the post-step \cf{PostGain}, both in sample.

\begin{table}[t]
  \caption{Stage ablation: change of the \vxcore{BenchRows}-problem mean (our meter) over each stage, runs 1--3 on the final
    benchmark. A problem without a stage keeps its score; cleanups are \$0; ``--'': stage off.}
  \label{tab:ablation}
  \centering\scriptsize
  \setlength{\tabcolsep}{4pt}
  \begin{tabular}{@{}llrrr@{}}
    \toprule
    Stage & Paid by & Run 1 & Run 2 & Run 3 \\
    \midrule
    Reference proof (score) & -- & \vxcore{SeedScore} & \vxcore{SeedScore} & \vxcore{SeedScore} \\
    CPU auto-brief + sessions to \$\cf{RunThreeRungOneUsd}, cleanup & Sol & \stg{One}{SolOneTwoFive}, \stg{One}{CleanOneTwoFive} & \stg{Two}{SolOneTwoFive}, \stg{Two}{CleanOneTwoFive} & \stg{Three}{SolOneTwoFive}, \stg{Three}{CleanOneTwoFive} \\
    Sessions to \$\cf{RunThreeRungTwoUsd}, cleanup & Terra & \stg{One}{TerraOneNineZero}, \stg{One}{CleanOneNineZero} & \stg{Two}{TerraOneNineZero}, \stg{Two}{CleanOneNineZero} & \stg{Three}{TerraOneNineZero}, \stg{Three}{CleanOneNineZero} \\
    Sessions to \$\cf{RunThreeLadderTopUsd}, cleanup & Terra & \stg{One}{TerraTwoSevenFive}, \stg{One}{CleanTwo} & \stg{Two}{TerraTwoSevenFive}, \stg{Two}{CleanTwo} & \stg{Three}{TerraTwoSevenFive}, \stg{Three}{CleanTwo} \\
    Review, repair, stack & Argon & \stg{One}{Review} & \stg{Two}{Review} & \stg{Three}{Review} \\
    Tail & Terra & \stg{One}{Tail} & \stg{Two}{Tail} & \stg{Three}{Tail} \\
    End stage + approach pick & -- & \stg{One}{Endstage} & \stg{Two}{Endstage} & \stg{Three}{Endstage} \\
    Style rewrite & Sol & \stg{One}{Stylerw} & \stg{Two}{Stylerw} & \stg{Three}{Stylerw} \\
    Dialect / hint / closer passes$^\ast$ & -- & -- & \stg{Two}{Dialect} / \stg{Two}{Endops} / \stg{Two}{DFour} & \stg{Three}{Dialect} / \stg{Three}{Endops} / \stg{Three}{DFour} \\
    \midrule
    End of run (our meter) & & \cf{RunOneStoredMean} & \cf{RunTwoStoredMean} & \cf{RunThreeStoredMean} \\
    Post-step, two passes$^\ast$ & -- & & & \cf{PostGain} \\
    Board & & \cf{RunOneBoardCombined}$^\dagger$ & \cf{RunTwoBoardCombined} & \textbf{\cf{UploadBoardCombined}} \\
    \bottomrule
  \end{tabular}\\[2pt]
  {\scriptsize $^\ast$Switched on or built after the release and chosen on the final problems, so in sample (Disclosures).
   $^\dagger$A probe upload of run~1.}
\end{table}

\subsection{Design decisions and evidence}
\label{sec:decisions}
Table~\ref{tab:decisions} lists the decisions behind the stages above with the comparisons that settled them. Most effects are small
next to the run-to-run spread, so several choices rest on mechanism (cost per turn, cap compliance) and on the absence of a measured loss.

\begin{table}[t]
  \caption{Design decisions. Unless marked, paired on the \vxcore{WarmupRows} public warm-up problems, one draw per arm; intervals:
    bootstrap 95\,\%. $\Delta$: score change per problem, points. ``Held-out'': a private unseen set, reported in
    aggregate. ``Final'': measured on final problems after the release (in sample).}
  \label{tab:decisions}
  \centering\scriptsize
  \setlength{\tabcolsep}{3pt}
  \begin{tabular}{@{}>{\raggedright}p{0.105\linewidth}p{0.70\linewidth}>{\raggedright\arraybackslash}p{0.145\linewidth}@{}}
    \toprule
    Decision & Evidence (alternatives tried in italics) & Choice \\
    \midrule
    Model per step &
    Terra vs \emph{Sol throughout} on later steps from the same \cf{DevTSevenRows} states (\usd{DevTSevenTerraUsd} vs
    \usd{DevTSevenSolUsd}): $\Delta=+\cf{DevTSevenGain}$ \ci{DevTSevenCiLo}{DevTSevenCiHi}; end to end, Sol then Terra vs Sol throughout
    $+\cf{DevArmsGain}$ \ci{DevArmsCiLo}{DevArmsCiHi}. \emph{Terra first} \cf{DevTerraFirstD} (\cf{DevTerraFirstRows} problems),
    \emph{Opus~5.5 second} \cf{DevOpusSecondD} (\cf{DevOpusSecondRows}; no interval). \emph{Sonnet~5.5 first}: \cf{DevSonnetHigh} vs Sol
    \cf{DevSonnetSol}, at \cf{DevSonnetCostRatioHigh}$\times$ Sol's cost/turn. \emph{Argon as main model} \cf{DevArgonMainA} at
    about \usd{DevArgonMainUsd} vs Sol \cf{DevArgonMainSol} at \usd{DevArgonMainSolUsd} (\cf{DevArgonMainRows} hard problems). &
    Sol to \$\cf{RunThreeRungOneUsd}, then Terra \\
    One budget ladder &
    \emph{A fresh Sol plan at \$\cf{RunThreeRungOneUsd}} $\Delta=\cf{DevReplanD}$ \ci{DevReplanCiLo}{DevReplanCiHi}
    (\cf{DevReplanRows} held-out problems, \usd{DevReplanUsdA} vs \usd{DevReplanUsdB}); \emph{twin sessions from one start}
    \cf{DevTwinsD} (\cf{DevTwinsRows} problems); \emph{a stuck problem sent to another model} \cf{DevRouteD} (\cf{DevRoutePairs}
    pairs); \emph{an early-stop rule} $+\cf{DevStopInGain}$ \cih{DevStopInLo}{DevStopInHi} in sample, \cf{DevStopOutGain}
    \ci{DevStopOutLo}{DevStopOutHi} held out. &
    one ladder, no stop rule \\
    Review &
    Multi-round $+\cf{DevReviewMultiSum}$ summed over \vxcore{WarmupRows} problems (\usd{DevReviewMultiUsd}) vs \emph{one round}
    $+\cf{DevReviewSingleSum}$ (\usd{DevReviewSingleUsd}); net of the end stage $+\cf{DevReviewNetGain}$
    \cih{DevReviewNetLo}{DevReviewNetHi} (\cf{DevReviewNetRowruns} problem-runs); repair + stack vs \emph{neither} $+\cf{DevRSixGain}$
    [+\cf{DevRSixCiLo}, +\cf{DevRSixCiHi}], \cf{DevRSixUp} up, \cf{DevRSixDown} down, \cf{DevRSixOneRowSharePct}\,\% of it on one
    problem. \emph{Other reviewers} (Kimi~K3, Astra): Appendix~\ref{app:experiments}. &
    Argon, \cf{RunThreeReviewRounds} rounds, repair + stack \\
    Late extras &
    Against \emph{none of them}: tail $+\cf{DevTailGain}$ \cih{DevTailCiLo}{DevTailCiHi} (\usd{DevTailSpendUsd} on \cf{DevTailRows}
    problems). Style rewrite $+\cf{DevStylerwHeldoutGain}$ \cih{DevStylerwHeldoutCiLo}{DevStylerwHeldoutCiHi} on
    \cf{DevStylerwHeldoutRows} held-out problems with \usd{DevStylerwHeldoutSpentLoUsd}--\cf{DevStylerwHeldoutSpentHiUsd} spent, but
    $+\cf{DevStylerwWarmupGain}$ on the warm-up with the budget left in the final runs. Approach pick $+\cf{DevApproachGain}$
    \cih{DevApproachCiLo}{DevApproachCiHi} out of sample (\cf{DevApproachSamples} samples, \cf{DevApproachDown} worse), below its
    pre-registered bar. &
    tail, style rewrite and, by the author's decision, approach pick \\
    End passes (post-release) &
    Final, \cf{DevBetsRows} problems, \emph{paid bets} (Terra at a stall, a planned restart, a profiler fix) vs control:
    $+\cf{DevBetsGain}$ \ci{DevBetsCiLo}{DevBetsCiHi}, so the bets were dropped. Final, \$0 replay on run~1's final proofs:
    $+\cf{DevReplayGain}$, of which $+\cf{DevReplayResearchRowsGain}$ on the problems the rules came from. &
    the three \$0 end passes \\
    Upload &
    Runs \cf{RunOneStoredMean} / \cf{RunTwoStoredMean} / \cf{RunThreeStoredMean} (\emph{run~1 or 2}); a \emph{per-problem union of
    runs} is not reproducible by the code. With \emph{\texttt{debug.skipKernelTC}} on \cf{DiscKskipRows} problems the board read
    \cf{KskipUploadBoard} vs \cf{NokskipVTwoUploadBoard} without it (same run, an earlier post-step); the option turns off Lean's kernel
    check. &
    run~3 + post-step, no skip: \cf{UploadBoardCombined} \\
    \bottomrule
  \end{tabular}
\end{table}

\subsection{Transfer from warm-up to the final benchmark}
\label{sec:transfer}
Most of those tests ran before the release; here is how they carried to the final benchmark. Run~1 ran code committed before the
release (\cf{RunOneCodeCommitUtc}~UTC); an earlier version was graded \cf{WarmupBoardMean} by the Arena on the warm-up. In our upload the
\cf{SharedRows} final problems that were public warm-up problems scored \cf{UploadStoredSharedMean}, the \cf{NewRows} new ones
\cf{UploadStoredNewMean}. On the shared problems run~2's upload scored \cf{RunTwoBoardMinusWarmupSharedPaired} against that warm-up
upload's grades (per-problem SD \cf{RunTwoBoardMinusWarmupSharedSd}), within noise; but \cf{SharedRowsNewerPin} of them list a newer first
toolchain on the final and our reference heartbeats differ on \cf{SharedRowsRefhbChanged}, so the grades lack a common reference. The new
problems' lower score mixes harder shapes with our development advantage on the shared ones; this comparison cannot separate them.

\subsection{What did not work}
\label{sec:negative}
Three kinds of effort did not pay inside the cap. \textbf{Prose knowledge.} In warm-up runs, sessions tried only \cf{DevLeadsTried} of the \cf{DevLeadsShown} library
lemmas their prompts suggested (\cf{DevLeadsPct}\,\%), and one card's literal edit failed to compile in \cf{DevCardFailed}
of \cf{DevCardMoves} forms tried at the state it targeted. \textbf{Rules built from the problems they score.} \cf{DevHbtokLiteral} of the
\cf{DevHbtokKinds} kinds of a heartbeat-rewrite module were literal templates of single warm-up proofs; it raised the warm-up from
\cf{DevHbtokWarmWithout} to \cf{DevHbtokWarmWith}, left a held-out set unchanged (\cf{DevHbtokHeldout}) and was deleted. Our
post-release passes deserve the same caution. \textbf{Plan diversity inside the cap.} Splits, twins, routes, re-plans
(Table~\ref{tab:decisions}) and a first session shown only statement and facts ($+\cf{DevBlindGain}$ \ci{DevBlindLo}{DevBlindHi}
vs a second control) did not beat the single ladder.

\section{Lessons, limitations and next steps}
\label{sec:lessons}
The results give lessons on four levels: Lean, the harness, the closed track and our open entry.
\paragraph{Lean-level lessons (both entries).}
(a)~\emph{Price every edit at the problem's own exchange rate.} $B/T$ ranges from \vxcore{BenchHbPerTokenMin} to
\vxcore{BenchHbPerTokenMax} heartbeats per token across the final problems, and in \exampleref{} \vxcore{ExHbDiff}
heartbeats outweigh \vxcore{ExTokensDiff} tokens; ranking by length alone, or by heartbeats alone, picks the wrong proof.
(b)~\emph{Length is what the lexer counts.} A dotted name or projection path (\texttt{e.typeCheck}, \texttt{t.2.1}) is a single
token, every bracket, comma and operator its own; several of our cheapest rewrites change no proof idea at all.
(c)~\emph{Check soundness with Lean, not a word filter.} A lakefile can turn \texttt{warn.sorry} off, so we force it on and audit
with \texttt{\#print axioms} outside \texttt{module} files. A \texttt{set\_option~...~in} chain above a declaration attaches to a
measuring command, not to the theorem, and can break reference readings.

\paragraph{Suggestions for the benchmark.} Ship the reference heartbeats, or the meter, with the final release; publish the token
counter (our copy matches every warm-up count but \vxcore{TokenizerFinalMatch} of the \vxcore{BenchRows} final ones); state which
toolchain $H$ is read on and how; and audit axioms in the evaluation worker, if it does not already, since a pattern filter sees
only the text of a proof.

\paragraph{One harness, two model classes.} The same harness design ran with an open model on rented GPUs
(\vxcore{BoardOpenCombined} on the open board) and with hosted models under the per-problem cap
(\vxcore{BoardClosedCombined} on the closed board). On our meter the closed upload is higher on
\vxcore{PairClosedHigherRows} of the \vxcore{BenchRows} problems and the open upload on \vxcore{PairOpenHigherRows}. We do not
attribute the \vxcore{BoardDiff}-point gap to the model class alone: the agent stages differ (models, budgets, sessions), as do
the post-step op lists and passes and each upload's disclosed choices.

\paragraph{From enterprise agents, and back.} Six of the harness's design principles come from enterprise agent systems from Vexorium: a
programmatic gate before any agent change counts; a spending guard that prices each call at its worst case before sending it; each
sub-task given the model that measured best per dollar in our own loop, not on a published benchmark; knowledge shipped as executed rules
rather than prompt text; every call recorded and every change reviewed by a person; and evidence that records its
limits~\citep{cheng2026evidence}. This entry applies them under a hard per-problem budget, within the limits disclosed below. We expect
them to help enterprise agents too, which we did not measure here.

\paragraph{What carried to the open entry.} Our two entries share one harness, and closed-track work reached the open entry as code and
settings, never as proofs: the paired bar (Section~\ref{sec:gate}); this entry's end-stage engine and generators, which the open run's
in-run polish used (its write-backs: $+\cf{OpenWritebackGainMeanPoints}$ on the open mean); and post-step ops learned from our closed
proofs. The \cf{XtLaneOps} ops named after closed-model research sessions add $+\cf{XtLaneOpsOpenMean}$ to the open upload's mean; the
\cf{XtLaneOpsInClosed} of them our passes ran add $+\cf{XtLaneOpsClosedMean}$ to ours, whose proofs were already polished (gain shares, not a removal test, in sample;
Appendix~\ref{app:devlessons} counts every op whose authors saw our closed proofs). Prose and model steps did not carry: with
GLM-5.3-Flash a closed-style rewrite and reviewer were accepted on \cf{OpenDevStyleAccepted} and \cf{OpenDevReviewAccepted} of
\cf{OpenDevStyleRows} problems, and a card written from comparing a closed model's sessions with Flash's showed no effect we
can tell from zero (\cf{OpenGlmCardDid} [\cf{OpenGlmCardDidLo}, \cf{OpenGlmCardDidHi}]). Work also flowed back: the dialect pass was built for the open
package first, the post-step code first ran on an open export, and the kernel-check skip, a construct the open model found first, was
tested in two of our uploads and withdrawn.

\paragraph{Against our open entry.} Seen from this side, the gap is one of score per output token: this run passed the open agent
stage's final mean after \cf{GaptokRunThreeReachOpenEndM}M output tokens, the open run needed \cf{GaptokOpenTotalM}M
(Figure~\ref{fig:gaptok}); prompts, tools and tokenizers differ, so the model's share cannot be split off (Appendix~\ref{app:devlessons}).

\paragraph{Lessons from the closed track.} (1)~\emph{Buy the plan early}: the first stage (auto-brief, then Sol) made
\cf{RunThreeFirstStageSharePct}--\cf{RunOneFirstStageSharePct}\,\% of the gain for \cf{RunThreeFirstStageSpendSharePct}\,\% of run~3's
spend. (2)~\emph{Execute, don't tell}: prompt advice was rarely acted on
(Section~\ref{sec:negative}); knowledge moved scores as gated passes, which also served our open entry, and, by hundredths, as worked
examples in the style rewrite (Table~\ref{tab:ablation}). (3)~\emph{Compare models in points per dollar under the real gate}: its worst-case reservation sets turns per
model (Sonnet~5.5's tokenizer counts the same Lean context \cf{DevSonnetTokenRatio}$\times$ longer than Sol's).
(4)~\emph{Single draws cannot resolve a few hundredths}: whole warm-up runs varied with an SD of \cf{DevWarmRunsSd}, above about
half of Table~\ref{tab:decisions}'s effects; we paired arms from shared states. (5)~\emph{Clean up between steps of one model}:
steps 2 and 3 both run Terra; between them, the \$\cf{RunThreeRungTwoUsd} checkpoint's cleanup found a better proof on
\cf{RunThreeFigCleanOneAdopted} of \cf{RunThreeFigCleanOneRows} problems and wrote it back, so step 3's sessions started from the
cleaned proof ($+\cf{RunThreeStageCleanOneNineZeroDelta}$ on the mean).

\paragraph{Limitations and next steps.} Post-release passes and the post-step were measured only in sample, and the cross-track
shares above are gain shares, not removal tests; the uploaded code ran once; most development comparisons are single draws on \vxcore{WarmupRows} problems. Next: measure those passes on unseen problems
and replace wall-clock limits with counted ones.

\paragraph{Disclosures.} \emph{Built or chosen after the release, on the final benchmark} (the code README's full list:
Appendix~\ref{app:disclosures}): (i)~the hint and closer passes: general rules we learned from uncapped research sessions of other closed
models on \cf{DiscResearchRows} final problems and from run~1's final proofs (no model was trained, no research proof is in any upload or
package, and the passes hold no problem names or proof text); (ii)~the choice of passes; (iii)~run~2's fixes, from run~1's records;
(iv)~the post-step's ops, learned the same way and from comparing our open and closed proofs of the final problems; one arm of those
research sessions also received material recovered from another entrant's public adapter, LeanZip~\citep{leanzip}
(\cf{LeanzipClosedOpsRan} of the ops our passes ran, \cf{LeanzipClosedOpsAdopted} adopted, $+\cf{LeanzipClosedGain}$ on the mean;
Appendix~\ref{app:ops}); (v)~run~3 on a second, same-type host. \emph{Also} (Appendix~\ref{app:disclosures}): a hand edit of
\cf{DiscRunOneEnveditRows} problems' environment files during run~1; two changes to the frozen code; style-rewrite example pairs of warm-up
theorems that are final problems, one shown to its own problem in \cf{DiscExemplarSelfrowPromptsSent} call; harness docstrings naming
library lemmas from our warm-up work for \cf{IsoDocstringNamedRows} final problems and hints for \cf{IsoDocstringHintRows} more, read by no session. \emph{Selection}: run~3 plus the post-step, chosen (with its passes and
ops) by our meter on the final problems over \cf{ClosedFinalUploads} scored uploads (\cf{ClosedFinalProbeUploads} probes, one an early
run-1 state uploaded as a lower placeholder; Appendix~\ref{app:uploads}); two carried \texttt{set\_option debug.skipKernelTC true in} on
\cf{DiscKskipRows} problems and were withdrawn. Of our \cf{WarmupPhaseUploads} warm-up uploads, \cf{WarmupPhaseUncappedUploads}, including
the last one shown (\cf{WarmupBoardLastCombined}), were development results not made under the cap from the reference proofs; that board
is not in the final ranking.
\paragraph{Models used only in development.} Besides each entry's run-time models (Section~\ref{sec:models}), development used
Claude Code (Claude Opus~5.5, Claude Sonnet~5.5) and Codex (GPT-6~Astra, GPT-6.1~Sol) as coding and research agents, and
GPT-6~Astra, Claude Opus~5.5, Fable~5.1 and Gemini~4 Argon in uncapped research sessions, some on final-benchmark problems; none of
their proofs is in an upload or a code package. Models measured and not adopted are in Appendix~\ref{app:experiments}.

\paragraph{Non-determinism.} Neither entry is bit-reproducible: model sampling and session scheduling vary between runs, and the
model-free stages depend on measured compile times. In one test, two runs of the same post-step pass on the same input (an earlier
op list, on the export of our second open-track run) changed the same \vxcore{RerunRowsChangedBoth} problems,
\vxcore{RerunRowsDiffer} of them to a different proof that passed the same gate; the means differed by \vxcore{RerunMeanDiff} points
(\vxcore{RerunMeanA} and \vxcore{RerunMeanB}). Section~\ref{sec:repro} gives our full runs' spread.

\paragraph{LLM assistance.} Claude Code wrote most of the harness code, supervised the runs, and assisted with this report; the
author takes responsibility for the content.

\vxbodyend
\begin{ack}
This work was supported by Azure credits from the Microsoft for Startups program, used for compute and for model API calls
during development. Google Cloud free-trial credits (Vertex AI) paid for part of the Gemini~4 Argon calls made on Vertex AI during development; we paid the rest. We received no other compute or API credits. Competing interests: none.
We thank Evan (\texttt{math-ai} on the Arena leaderboard) for a Lean metaprogramming idea.

\end{ack}

\vxreferences

\appendix

\section{Per-problem spend and results}
\label{app:perproblem}
\begin{figure}[h]
  \centering
  \input{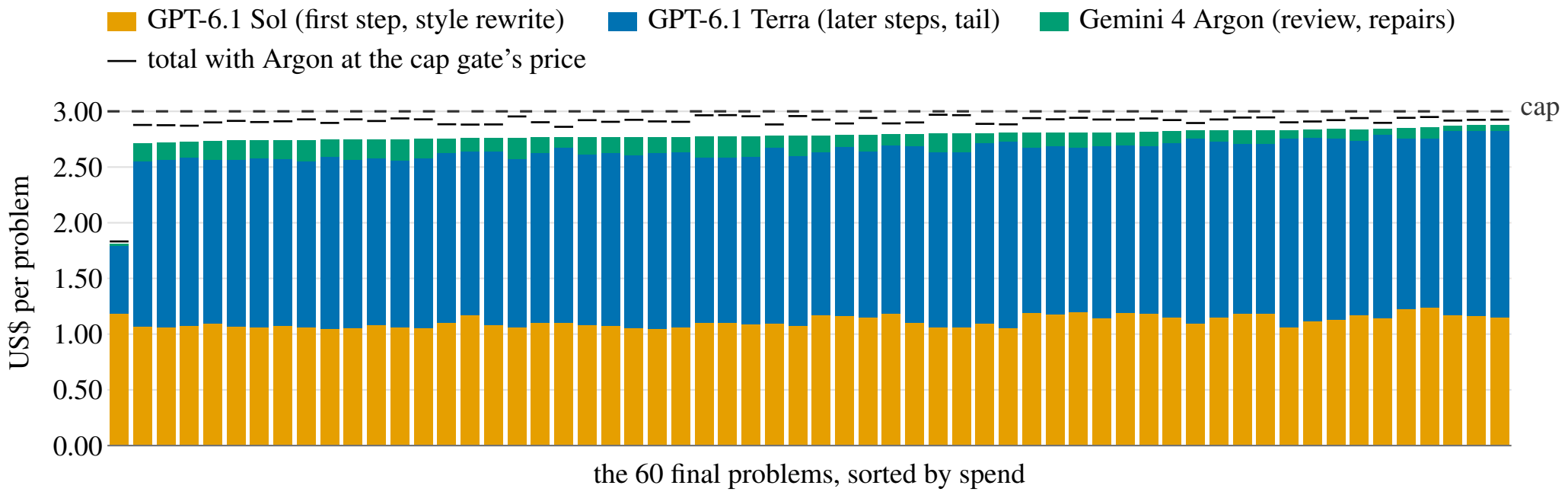}
  \caption{Per-problem spend of the uploaded run at list prices, stacked by model and sorted. A black tick marks each problem's
    total with Argon at the cap gate's price; the dashed line is the cap.}
  \label{fig:spend}
\end{figure}

Table~\ref{tab:perproblem} lists every problem of the uploaded run: its spend at list prices by model (all calls, as in
Table~\ref{tab:budget}; the Sol column includes the style rewrite), its total with Argon at the cap gate's price, and the upload's
score on our meter with its $L$ and $H$ ($Z=\vxcore{FullZ}$ on every problem). $L$ and $H$ are computed from the stored tokens and heartbeats
against the references. The score is the pipeline's paired reading at adoption: on \ClosedFigScoreDiffRows{} problems it differs from
$(L+H+Z)/3$ by at most \ClosedFigScoreDiffMax{}. $^\dagger$: one of the \vxcore{WarmupSharedRows} public warm-up problems. Figure~\ref{fig:spend}
shows the spend against the cap: \cf{RunThreeRowsAboveTwoNineCapPrice} problems end above \usd{RunThreeNearCapThresholdUsd} at the cap
price, because the review, which the cap gate prices at Argon's doubled rate, spends most of the room the budget steps leave. The rows are generated from the run records by
\texttt{report/figures/make\_closed\_figs.py}, which checks the totals against the fact sheet. Figure~\ref{fig:curve} gives the mean score
against mean spend after each stage of runs 1--3. In Figure~\ref{fig:lib}, PutnamBench is high on $L$ and $H$ (mean
\cf{LibPutnambenchUploadStoredMean}) and PhysLib (\cf{LibPhyslibUploadStoredMean}) holds the problems whose heartbeats barely moved.

\begin{figure}[h]
  \centering
  \begin{minipage}[b]{0.565\linewidth}
    \centering
    \begin{tikzpicture}[font=\scriptsize]
\draw[black!12] (0.256,0) -- (0.256,2.600);
\node[below,inner sep=1pt] at (0.256,0) {1.00};
\draw[black!12] (1.535,0) -- (1.535,2.600);
\node[below,inner sep=1pt] at (1.535,0) {1.50};
\draw[black!12] (2.814,0) -- (2.814,2.600);
\node[below,inner sep=1pt] at (2.814,0) {2.00};
\draw[black!12] (4.093,0) -- (4.093,2.600);
\node[below,inner sep=1pt] at (4.093,0) {2.50};
\draw[black!12] (5.372,0) -- (5.372,2.600);
\node[below,inner sep=1pt] at (5.372,0) {3.00};
\draw[black!12] (0,0.163) -- (5.500,0.163);
\node[left,inner sep=1pt] at (0,0.163) {81};
\draw[black!12] (0,0.569) -- (5.500,0.569);
\node[left,inner sep=1pt] at (0,0.569) {82};
\draw[black!12] (0,0.975) -- (5.500,0.975);
\node[left,inner sep=1pt] at (0,0.975) {83};
\draw[black!12] (0,1.381) -- (5.500,1.381);
\node[left,inner sep=1pt] at (0,1.381) {84};
\draw[black!12] (0,1.788) -- (5.500,1.788);
\node[left,inner sep=1pt] at (0,1.788) {85};
\draw[black!12] (0,2.194) -- (5.500,2.194);
\node[left,inner sep=1pt] at (0,2.194) {86};
\draw[black!55] (0,0) rectangle (5.500,2.600);
\draw[black!70,dashed] (5.372,0) -- (5.372,2.600) node[pos=0.5,rotate=90,above,inner sep=1pt] {cap};
\node[below=8pt] at (2.750,0) {mean spend per problem (US\$, list price)};
\node[rotate=90] at (-0.62,1.300) {60-problem mean score (our meter)};
\draw[vxGray, densely dotted, line width=0.6pt] (0.471,0.411) -- (0.471,0.565) -- (2.168,1.300) -- (2.168,1.340) -- (4.343,1.834) -- (4.343,1.865) -- (4.649,1.928) -- (4.776,1.931) -- (4.776,2.064) -- (4.858,2.076);
\draw[vxGray, dashed, line width=0.6pt] (0.472,0.351) -- (0.472,0.524) -- (2.140,1.215) -- (2.140,1.254) -- (4.283,1.766) -- (4.283,1.796) -- (4.607,1.868) -- (4.730,1.869) -- (4.730,2.018) -- (4.783,2.018) -- (4.783,2.039) -- (4.783,2.039) -- (4.783,2.106);
\draw[black, line width=0.9pt] (0.458,0.218) -- (0.458,0.397) -- (2.137,1.284) -- (2.137,1.315) -- (4.275,1.847) -- (4.275,1.879) -- (4.601,1.941) -- (4.695,1.943) -- (4.695,2.082) -- (4.789,2.091) -- (4.789,2.129) -- (4.789,2.129) -- (4.789,2.197);
\fill[black] (0.458,0.218) circle (0.9pt);
\fill[black] (0.458,0.397) circle (0.9pt);
\fill[black] (2.137,1.284) circle (0.9pt);
\fill[black] (2.137,1.315) circle (0.9pt);
\fill[black] (4.275,1.847) circle (0.9pt);
\fill[black] (4.275,1.879) circle (0.9pt);
\fill[black] (4.601,1.941) circle (0.9pt);
\fill[black] (4.695,1.943) circle (0.9pt);
\fill[black] (4.695,2.082) circle (0.9pt);
\fill[black] (4.789,2.091) circle (0.9pt);
\fill[black] (4.789,2.129) circle (0.9pt);
\fill[black] (4.789,2.129) circle (0.9pt);
\fill[black] (4.789,2.197) circle (0.9pt);
\draw[vxVermillion, line width=0.9pt] (4.789,2.197) -- (4.789,2.295) -- (4.789,2.319);
\fill[vxVermillion] (4.789,2.295) circle (1.1pt);
\fill[vxVermillion] (4.789,2.319) circle (1.1pt);
\node[star,star points=5,star point ratio=2.2,fill=vxVermillion,inner sep=0pt,minimum size=5pt] at (4.969,2.320) {};
\node[right,inner sep=1pt] at (0.508,0.284) {Sol sessions (from the reference)};
\node[anchor=north west,inner sep=1pt] at (3.453,1.503) {Terra steps};
\node[left,inner sep=1pt] at (4.502,2.112) {review, tail, \$0 end stages};
\node[left,inner sep=1pt] at (4.709,2.389) {\textcolor{vxVermillion}{post-step, board}};
\draw[black, line width=0.9pt] (0.12,2.38) -- (0.47,2.38) node[right,inner sep=2pt] {run 3 (uploaded)};
\draw[vxGray, dashed, line width=0.6pt] (0.12,2.13) -- (0.47,2.13) node[right,inner sep=2pt] {run 2};
\draw[vxGray, densely dotted, line width=0.6pt] (0.12,1.88) -- (0.47,1.88) node[right,inner sep=2pt] {run 1};
\end{tikzpicture}
    \caption{Mean score vs mean spend per problem after each stage (vertical steps: \$0 stages; reference proofs,
      \vxcore{SeedScore} at \$0, off the chart). ``Post-step, board'': run~3's post-step passes; board score (star).}
    \label{fig:curve}
  \end{minipage}\hfill
  \begin{minipage}[b]{0.41\linewidth}
    \centering
    \begin{tikzpicture}[font=\scriptsize]
\draw[black!12] (0.000,0) -- (0.000,2.600);
\draw[black!12] (0,0.000) -- (2.600,0.000);
\node[below,inner sep=1pt] at (0.000,0) {0};
\node[left,inner sep=1pt] at (0,0.000) {0};
\draw[black!12] (0.650,0) -- (0.650,2.600);
\draw[black!12] (0,0.650) -- (2.600,0.650);
\node[below,inner sep=1pt] at (0.650,0) {25};
\node[left,inner sep=1pt] at (0,0.650) {25};
\draw[black!12] (1.300,0) -- (1.300,2.600);
\draw[black!12] (0,1.300) -- (2.600,1.300);
\node[below,inner sep=1pt] at (1.300,0) {50};
\node[left,inner sep=1pt] at (0,1.300) {50};
\draw[black!12] (1.950,0) -- (1.950,2.600);
\draw[black!12] (0,1.950) -- (2.600,1.950);
\node[below,inner sep=1pt] at (1.950,0) {75};
\node[left,inner sep=1pt] at (0,1.950) {75};
\draw[black!12] (2.600,0) -- (2.600,2.600);
\draw[black!12] (0,2.600) -- (2.600,2.600);
\node[below,inner sep=1pt] at (2.600,0) {100};
\node[left,inner sep=1pt] at (0,2.600) {100};
\draw[black!55] (0,0) rectangle (2.600,2.600);
\draw[black!40, dashed] (0.260,2.600) -- (2.600,0.260) node[pos=0,above,inner sep=1pt,black!60] {70};
\draw[black!40, dashed] (1.040,2.600) -- (2.600,1.040) node[pos=0,above,inner sep=1pt,black!60] {80};
\draw[black!40, dashed] (1.820,2.600) -- (2.600,1.820) node[pos=0,above,inner sep=1pt,black!60] {90};
\node[below=8pt] at (1.300,0) {$H$ (heartbeat reduction, \%)};
\node[rotate=90] at (-0.72,1.300) {$L$ (length reduction, \%)};
\draw[only marks,mark=*,mark size=1.7pt,color=vxPurple,mark options={draw=white,line width=0.2pt}] plot coordinates {(2.544,2.456) (2.433,2.536) (2.489,2.465) (2.487,2.448) (2.582,2.563) (2.545,2.562) (2.567,2.499) (2.450,2.420) (2.544,2.442) (2.591,2.460)};
\draw[only marks,mark=square*,mark size=1.7pt,color=vxBlue,mark options={draw=white,line width=0.2pt}] plot coordinates {(2.240,2.427) (2.319,2.465) (2.289,2.280) (2.531,2.052) (2.204,1.802) (2.504,2.406) (2.544,2.175) (2.306,2.160) (2.369,2.100) (2.452,1.769)};
\draw[only marks,mark=triangle*,mark size=1.7pt,color=vxGreen,mark options={draw=white,line width=0.2pt}] plot coordinates {(2.466,1.776) (1.868,1.542) (2.212,2.108) (2.411,2.225) (2.355,2.289) (1.694,1.744) (2.140,2.230) (1.776,1.539) (2.368,2.453) (1.378,2.084)};
\draw[only marks,mark=diamond*,mark size=1.7pt,color=vxSky,mark options={draw=white,line width=0.2pt}] plot coordinates {(2.516,2.252) (2.124,2.034) (1.980,2.110) (2.283,2.365) (1.180,1.769) (2.025,1.908) (2.169,2.011) (1.502,1.318) (2.522,1.830) (2.497,2.046)};
\draw[only marks,mark=pentagon*,mark size=1.7pt,color=vxOrange,mark options={draw=white,line width=0.2pt}] plot coordinates {(2.367,2.160) (1.844,1.635) (2.413,1.896) (2.209,1.716) (2.390,1.123) (2.390,1.323) (2.537,2.035) (1.870,1.608) (2.494,1.671) (2.318,1.543)};
\draw[only marks,mark=x,mark size=1.7pt,color=vxVermillion,mark options={line width=0.8pt}] plot coordinates {(0.000,0.000) (-0.001,1.300) (-0.001,1.300) (1.448,1.954) (1.298,1.804) (2.152,2.094) (2.168,2.038) (2.312,2.313) (2.541,2.389) (2.550,2.169)};
\draw[only marks,mark=*,mark size=1.7pt,color=vxPurple] plot coordinates {(2.90,2.40)};
\node[right,inner sep=2pt] at (2.95,2.40) {PutnamBench};
\draw[only marks,mark=square*,mark size=1.7pt,color=vxBlue] plot coordinates {(2.90,2.10)};
\node[right,inner sep=2pt] at (2.95,2.10) {ArkLib};
\draw[only marks,mark=triangle*,mark size=1.7pt,color=vxGreen] plot coordinates {(2.90,1.80)};
\node[right,inner sep=2pt] at (2.95,1.80) {Strata};
\draw[only marks,mark=diamond*,mark size=1.7pt,color=vxSky] plot coordinates {(2.90,1.50)};
\node[right,inner sep=2pt] at (2.95,1.50) {LeanPool};
\draw[only marks,mark=pentagon*,mark size=1.7pt,color=vxOrange] plot coordinates {(2.90,1.20)};
\node[right,inner sep=2pt] at (2.95,1.20) {CSLib};
\draw[only marks,mark=x,mark size=1.7pt,color=vxVermillion,mark options={line width=0.8pt}] plot coordinates {(2.90,0.90)};
\node[right,inner sep=2pt] at (2.95,0.90) {PhysLib};
\end{tikzpicture}
    \caption{The upload per problem, by library: $L$ against $H$ (our meter). Dashed: equal score ($Z=\vxcore{FullZ}$).}
    \label{fig:lib}
  \end{minipage}
\end{figure}

{\scriptsize
\setlength{\tabcolsep}{3pt}
\begin{longtable}{@{}p{0.38\linewidth}rrrrrrrr@{}}
  \caption{Per-problem spend (US\$, list prices) and result of the uploaded run.}\label{tab:perproblem}\\
  \toprule
  Problem & Sol & Terra & Argon & Total & At cap price & Score & $L$ & $H$ \\
  \endfirsthead
  \toprule
  Problem & Sol & Terra & Argon & Total & At cap price & Score & $L$ & $H$ \\
  \endhead
  \ClosedPerProblemRows
  \bottomrule
\end{longtable}
}

\section{Experiment log}
\label{app:experiments}
Table~\ref{tab:explog} lists the development tests behind the closed pipeline, one line each, in date order within four phases. The
numbers of the tests that decided the design are in Table~\ref{tab:decisions}. ``Warm-up'': the public warm-up problems;
``held-out'': private sets of unseen problems, reported only as aggregates; ``final'': the final benchmark after its release (in
sample). Dates are the start of each test (month-day, from our experiment catalog).

\paragraph{Models measured and not adopted.} Claude Opus~5.5 (as the capped engine in Claude Code, as the second-step model, at a
stall), Claude Sonnet~5.5 (first step), Gemini~4 Argon (as the main model; it stays the reviewer), GPT-6~Astra (as the reviewer, and
uncapped research), Fable~5.1 (uncapped research), Kimi~K3 (reviewer and deeper sessions), DeepSeek~V4.1 Flash (an executor of review
edits), GPT-6.1 Terra as the first-step model and GPT-6.1 Sol at extra-high effort.

{\scriptsize
\setlength{\tabcolsep}{3pt}
\begin{longtable}{@{}p{0.06\linewidth}p{0.13\linewidth}p{0.40\linewidth}p{0.36\linewidth}@{}}
  \caption{Experiment log of the closed track.}\label{tab:explog}\\
  \toprule
  Start & Data & Tested & Outcome \\
  \endfirsthead
  \toprule
  Start & Data & Tested & Outcome \\
  \endhead
  \ClosedExpLogRows
  \bottomrule
\end{longtable}
}

\section{Development lessons and what crossed between our entries}
\label{app:devlessons}
This appendix gathers what the closed track learned in development and what crossed between our two entries, which were built side by
side on one harness. Sets and readings are as in Appendix~\ref{app:experiments}.

\paragraph{What crossed, counted.} Of the \cf{XtOpenPostOps} ops of our open upload's post-step, \cf{XtOpsClosedProofs} were written by
research or coding agents that studied our closed proofs of the same final problems (\cf{XtOpsFromClosedLine} of them copied from this
entry's end passes), \cf{XtOpsClosedStats} by studies that used only aggregate statistics of our closed proofs, and \cf{XtOpsNoClosed} have
no closed origin. The first group carries \cf{XtOpsClosedProofsPointsPct}\,\% of the open post-step's measured points
($+\cf{XtOpsClosedProofsMean}$ on the open mean). These are gain shares: each adopted step is credited to the op that made it, and other ops
might have made some of those steps had it been missing; all of it is in sample. In their first two passes, at about the same compile count
(\cf{OpenGapFirstPassesOpenCompiles} and \cf{OpenGapClosedPostCompiles}), the open post-step gained $+\cf{OpenGapFirstPassesOpenNoKskip}$
without the kernel-check skip and ours $+\cf{OpenGapFirstPassesClosed}$: the same kind of rules pay most on proofs a weaker model left
unpolished. The open package holds no closed proof text, and the closed archive only the style rewrite's example pairs
(Appendix~\ref{app:disclosures}); the ops hold no problem names.

\paragraph{What the task asks of a model.} Refactoring a working proof has no natural end and the compiler checks every attempt, so
the uploaded run spent its dollars on many checked attempts rather than depth in one answer: \cf{OpenGapTurnsPerRowClosed} turns per
problem (Sol \cf{OpenLoopTurnsPerRowSol}, Terra \cf{OpenLoopTurnsPerRowTerra}), with \cf{OpenLoopToolCallsPerRowSol} and
\cf{OpenLoopToolCallsPerRowTerra} tool calls, of them \cf{OpenLoopCheckCallsPerRowSol} and \cf{OpenLoopCheckCallsPerRowTerra} compile
calls; \cf{OpenLoopCompileFailPctSol}\,\% (Sol) and \cf{OpenLoopCompileFailPctTerra}\,\% (Terra) of the candidates failed to compile and
\cf{OpenLoopUnusablePctSol}\,\% of the calls were unusable. Deeper thinking per turn bought half the turns and no measured gain
(Section~\ref{sec:agent}).

\begin{figure}[h]
  \centering
  \input{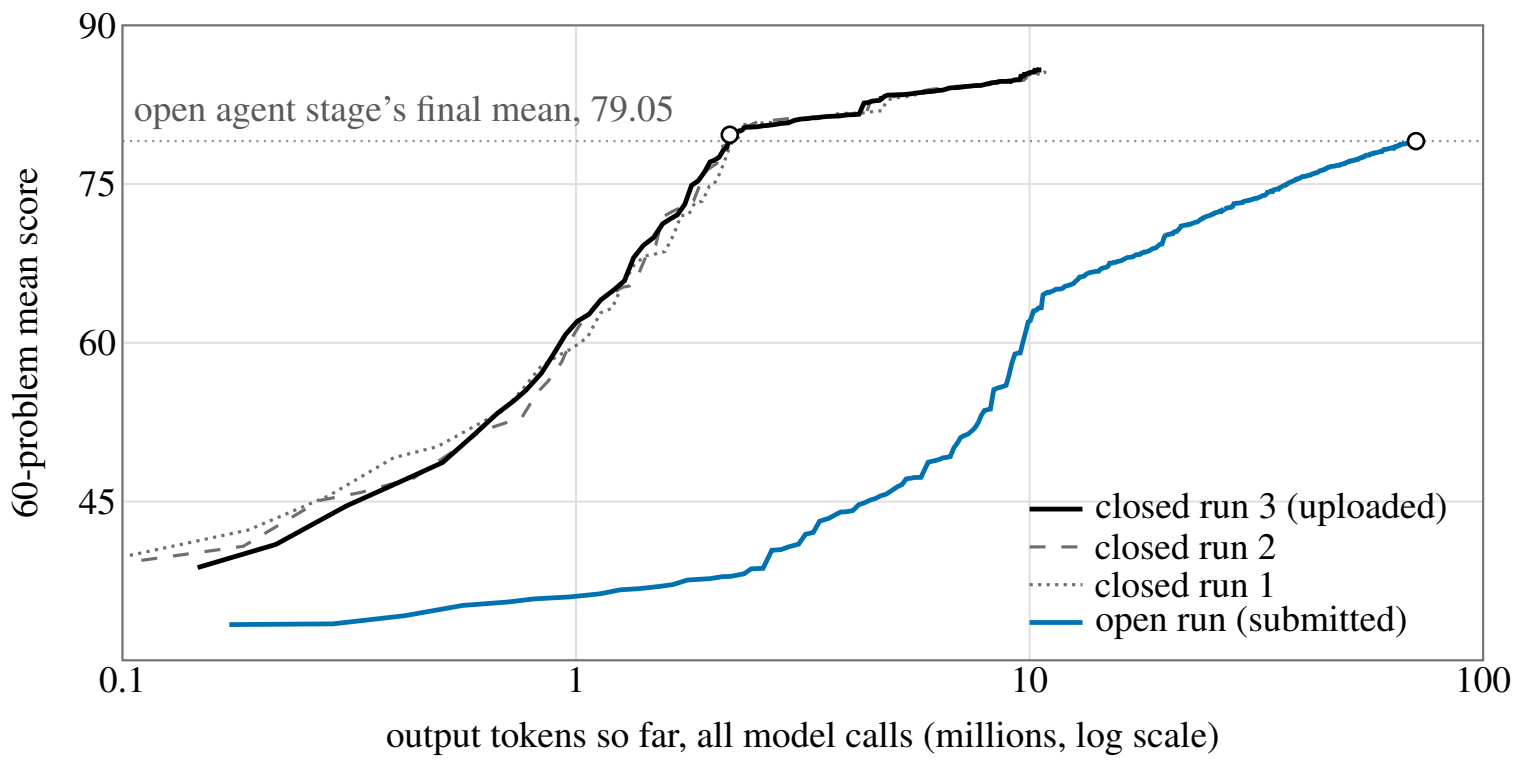}
  \caption{Mean score (our meter) against cumulative output tokens: our closed runs 1--3 and the submitted open run (log scale). Each
    model counts tokens with its own tokenizer; the closed curves start above the reference mostly with model turns, partly with the
    auto-brief's starts. The open curve ends with its agent stage, the closed ones with their runs' end stages.}
  \label{fig:gaptok}
\end{figure}

\paragraph{Score per output token.} After this run's whole \cf{GaptokRunThreeTotalM}M output tokens the open run stood at
\cf{GaptokOpenAtRunThreeTotal}; it needed \cf{GaptokReachRatio} times this run's tokens to reach the mean this run had passed after
\cf{GaptokRunThreeReachOpenEndM}M (Figure~\ref{fig:gaptok}). This pipeline passed the open upload's board score
(\cf{OpenBoardCombined}) between its first budget step (\cf{OpenGapClosedStageOne} at a mean US\$\cf{OpenGapClosedStageOneUsd}) and its
second (\cf{OpenGapClosedStageTwo} at US\$\cf{OpenGapClosedStageTwoUsd}). Prompts, tools and the auto-brief differ, and each model counts
tokens with its own tokenizer, so the model's share cannot be split off.

\paragraph{Lessons from development.} Each item is a development comparison, mostly single draws, unless marked otherwise.
\begin{list}{\textbullet}{\setlength{\leftmargin}{0.9em}\setlength{\labelwidth}{0.6em}\setlength{\labelsep}{0.3em}%
  \setlength{\itemsep}{1pt}\setlength{\parsep}{0pt}\setlength{\topsep}{2pt}}
\item \emph{Match models to roles under the real gate.} On \cf{DlDepthRows} hard warm-up problems, one US\$\cf{DlDepthCapUsd} session
  each from the same proof, Sol at extra-high effort gained \cf{DlDepthSolPpd} points per dollar, Claude Opus~5.5 \cf{DlDepthOpusPpd} and
  DeepSeek V4.1 Flash \cf{DlDepthFlashPpd}. The gate's worst-case reserve sets a model's turns: Opus got
  \cf{DlDepthOpusTurnsLo}--\cf{DlDepthOpusTurnsHi} and stopped at US\$\cf{DlDepthOpusStopLoUsd}--\cf{DlDepthOpusStopHiUsd}. None of
  \cf{DlArchitectModels} models asked without tools for a new proof skeleton gave a usable one. In review, breadth beat precision, since
  compiles are free: Argon proposed \cf{DlReviewArgonEdits} edits (\cf{DlReviewArgonCompilePct}\,\% compiled) for $+\cf{DevReviewMultiSum}$
  summed over \vxcore{WarmupRows} problems, GPT-6 Astra \cf{DlReviewAstraEdits} (\cf{DlReviewAstraCompilePct}\,\%) for
  $+\cf{DlReviewAstraMultiSum}$. An OpenAI reviewer in the same frame reached about \cf{DlReviewOpenaiSharePct}\,\% of Argon's gain: most
  of the value is the format (exact local edits, several rounds, each compiled).
\item \emph{How the pre-release rules were made and gated.} The \$0 rule families were learned from winning edits and heartbeat profiles
  of our development runs on the warm-up (uncapped ones with closed models included), written without problem names, and audited: every
  token a rule inserts must come from the proof, the rule's text or a table generated from Lean. Each family was then replayed on stored
  final proofs of \cf{DlOpsHeldoutRows} held-out problems, none used in design. Micro-edit and heartbeat-swap families gained
  $+\cf{DlOpsWarmGain}$ a problem on the warm-up and $+\cf{DlOpsHeldoutGain}$ held out (\cf{DlOpsHeldoutUp} improved), the heartbeat tool
  $+\cf{DlHbtoolWarmGain}$ and $+\cf{DlHbtoolHeldoutGain}$ (\cf{DlHbtoolHeldoutUp}); both were adopted. A simp-merging family improved
  \cf{DlOpsSimpHeldoutUp} of them and stayed off. A structure op had one look at a second held-out set: it fired on \cf{DlStructopFired} of
  \cf{DlStructopRows}, $+\cf{DlStructopGain}$ a problem, none lower in $Z$. Rebuild rules tuned to the best warm-up proofs added
  $+\cf{DlMinedWarmGain}$ to the warm-up mean but fired on \cf{DlMinedFiresA} of \cf{DlMinedRowsA} and \cf{DlMinedFiresB} of
  \cf{DlMinedRowsB} held-out problems; they were held back.
\item \emph{The records caught defects that tests missed.} Auto-briefs built in containers skipped the heartbeat tool silently
  (\cf{DlBriefSkip} of \cf{DlBriefSkipOf}); the fix raised the first session's start by $+\cf{DlBriefFixStartGain}$ a warm-up problem
  (\cf{DlBriefFixUp} better, \cf{DlBriefFixDown} worse) and $+\cf{DlBriefFixNwGain}$ a problem on a set of other development
  problems (\cf{DlBriefFixNwRows} of them briefed before a timeout, the rest counted as unchanged). Runaway tool calls ran to the output cap but were reported as tool calls: \cf{DlJunkTurns} such turns took
  \cf{DlJunkSpendPct}\,\% of one run's spend; a detector caught \cf{DlJunkCaught} of \cf{DlJunkCaughtOf} with \cf{DlJunkFalse} false hits
  in \cf{DlJunkLegit} calls. Provider overloads were booked at worst case (US\$\cf{DlFivexxUsd} in one run). Two reviewers found final
  problems with warm-up names scored against warm-up reference heartbeats (up to \cf{DlCalibPct}\,\% higher); an environment fix took
  Lean's heartbeat timeout for a process timeout (about \cf{DlTimeoutMeanLoss} on the mean in a rerun); on PhysLib, which turns off the
  \texttt{sorry} warning, a proof resting on \texttt{sorryAx} posed as a $+\cf{DlSorryFakeGain}$ gain. All are fixed in the uploaded code.
\item \emph{The author's review changed decisions.} The author challenged the reviewer's place; the study that followed found its value
  on held-out problems (below), so it stayed. The author kept the review's repair call rather than a stack-only variant: in its test the
  repair call made \cf{DevRSixOneRowSharePct}\,\% of the gain and the stack without repairs \cf{DlStackOnlyPct}\,\%, so the choice
  kept \cf{DlRepairKeptPct}\,\%. A model that would stop sessions early
  was tried before it was judged (US\$\cf{DlDecisionUsd}): after one turn it ranked \cf{DlDecisionSessions} sessions at AUC
  \cf{DlDecisionAuc}, near chance, and stopping on it would have changed the summed score by \cf{DlDecisionStopLoss}. The author declined
  a fourth full run worth an expected $+\cf{DlRunfourLo}$ to $+\cf{DlRunfourHi}$.
\item \emph{The plateau belongs to the method; selection is what a cap cannot buy.} Uncapped on the warm-up, Claude Code with Opus~5.5
  and Codex with GPT-6 Astra, resumed over passes, plateaued at \cf{DlUncappedOpus} and \cf{DlUncappedAstra}, and the per-problem
  best of all our lines at \cf{DlUncappedBestPlateau}. It reached \cf{DlUncappedBestEnd} through steps outside a single trajectory, chiefly a review
  naming exact edits with a budgeted compile of them, plus a model-free normalizer: the forms our capped review and \$0 stages took.
  Under the cap, the gap from our warm-up runs to the best uncapped proofs was \cf{DlGapVariancePct}\,\% variance across draws and
  \cf{DlGapCapabilityPct}\,\% moves no capped run found; the best of \cf{DlBestofLoK} to \cf{DlBestofHiK} whole runs would add
  $+\cf{DlBestofLoGain}$ to $+\cf{DlBestofHiGain}$ at that many times the budget. In the frozen pipeline \cf{DlDeadMoneyPct}\,\% of
  session money bought no verified gain; no tested signal flagged it.
\item \emph{Late stages pay where proofs are unpolished, so warm-up tests understate them.} Net of the end stage, the review gained
  $+\cf{DlReviewNetWarmChangedGain}$ per problem it changed on warm-up runs but $+\cf{DlReviewNetHeldoutGain}$
  \cih{DlReviewNetHeldoutLo}{DlReviewNetHeldoutHi} on the \cf{DlReviewNetHeldoutRows} held-out problems it changed
  (\cf{DlReviewNetHeldoutUp} better, \cf{DlReviewNetHeldoutDown} worse; $+\cf{DlReviewNetHeldoutNodeadGain}$ without a dead session), and \cf{DlReviewPpdLow} points per dollar below a score of \cf{DlBandEdgeLo}. On run~1's final proofs (in sample) the end stage gained
  $+\cf{DlEndBandLow}$ a problem below \cf{DlBandEdgeLo}, $+\cf{DlEndBandMid}$ up to \cf{DlBandEdgeHi} and $+\cf{DlEndBandHigh}$ above
  (\cf{DlEndBandLowRows}, \cf{DlEndBandMidRows} and \cf{DlEndBandHighRows} problems); the style rewrite splits the same way
  (Table~\ref{tab:decisions}).
\item \emph{Agent studies, checked by skeptics.} Most decisions rested on read-only agent studies of our records (one read
  \cf{DlRecordsProofs} verified proofs of \cf{DlRecordsRuns} runs), many with a skeptic or verifier pass. First numbers often moved: a claimed
  scaffold share of at least \cf{DlScaffoldClaimPct}\,\% of a gain fell to \cf{DlScaffoldLoPct}--\cf{DlScaffoldHiPct}\,\% policy; a
  held-out review gain of $+\cf{DlReviewRetiredGain}$ a problem was retired (it predated the end stage; a dead session drove it); a
  reviewer's own forecast fell \cf{DlReviewerEstimateOff} points short; a gap of \cf{DlFirsthalfGap} between serving routes was about
  \cf{DlFirsthalfNoisePct}\,\% noise.
\item \emph{Compare against draw populations, not one control.} On \vxcore{WarmupRows} warm-up problems an arm difference has a standard
  error of \cf{DlArmSe} (even a forked pair on the problems without a structural jump has only about \cf{DlPowerPct}\,\% power
  against a true \cf{DlPowerEffect}); confirming $+\cf{DlPowerSmallEffect}$ a
  problem needs about \cf{DlPowerRowsLo}--\cf{DlPowerRowsHi} paired problems. Controls can be lucky: Sol at extra-high effort read
  \cf{OpenSolXhighPaired} [\cf{OpenSolXhighBestLo},\,\cf{OpenSolXhighBestHi}] against the best of \cf{OpenSolMediumDraws} medium draws,
  but $+\cf{DlXhighLatestGain}$ \ci{DlXhighLatestLo}{DlXhighLatestHi} against the latest \cf{DlXhighLatestN}.
\item \emph{Protect the run before chasing gain; enforce isolation.} A failure-path audit found \cf{DlAuditHarms} harms,
  \cf{DlAuditCritical} of them critical: a single failing problem could end the run with no output, nothing wrote a submission until the
  end, state was not written atomically, and concurrent sessions or a trusted provider cost field could push a problem over the
  cap. Replaying \cf{DlAuditTurns} recorded turns found \cf{DlAuditGateViolations} breaches of the gate's bound. The uploaded code writes
  the submission from the start and on every exit. Isolation was tightened after development tests broke it; none was an uploaded
  run. On \cf{IsoPilotDate} a warm-up pilot session (Sol through a vendor command-line tool we then dropped) copied another
  development line's proof of a warm-up problem that is not a final problem; we removed the pilot's records. On \cf{IsoSnapshotDate}
  a model-versus-scaffold test on \cf{IsoSnapshotRows} warm-up problems that are also final problems (uncapped Claude Opus~5.5 and
  GPT-6 Astra) let both arms read our pooled proofs, though only one arm was meant to: we withdrew its reading and reran it without
  the pool, but one of its proofs and a later descendant went into \cf{IsoSnapshotUploads} of our uncapped warm-up-board uploads
  (Disclosures), none into a final upload. On \cf{IsoGlmDate} an open-model test in the development host's shell read a closed-line
  candidate and was stopped. On \cf{IsoNotesDate}, after the release, \cf{IsoNotesTests} capped tests in that shell read a shared
  notes file and a compile-scratch file holding our research proofs of \cf{IsoNotesRows} final problems; their results were
  retracted or discarded. No uploaded run could read these files: our closed runs ran on separate cloud VMs, each session in a
  container with a whitelisted checkout of the code, its shard's work and the environments, and in their \cf{IsoClosedSessions}
  sessions (\cf{IsoClosedShellCalls} shell calls) no command or file tool reached another line's or a research folder, run records,
  pooled proofs, the harness code or another problem's workspace (in run~1, \cf{IsoRunOneScratchSessions} session printed a compile
  file of another problem of its shard; none of it reached its proofs). Every measurement test now runs in a container (our uncapped
  research sessions stayed on the host), and every run first passes a leak scan of model-facing text; the scan skips docstrings and comments, which the sessions could read but did not
  (Disclosures).
\end{list}

\section{Environment pins}
\label{app:pins}
{\small
\begin{longtable}{@{}p{0.27\linewidth}p{0.70\linewidth}@{}}
  \caption{Environment pins of the uploaded run.}\label{tab:pins}\\
  \toprule
  Item & Pin \\
  \midrule
  Code repository & \url{https://github.com/bobbercheng/lean-refactor-closed} \\[3pt]
  Code archive there & \texttt{final/} at tag \texttt{closed-final-sub41-20261104} \\
  Code archive (sha256) & \texttt{\scriptsize\cf{CodeZipShaTwoFiveSix}} \\
  Run~3 code (the archive) & commit \texttt{\cf{RunThreeCodeRef}}, \cf{RunThreeCodeCommitUtc}~UTC \\
  Run~2 / run~1 code & \texttt{\cf{RunTwoCodeRef}} (\cf{RunTwoCodeCommitUtc}~UTC) / \texttt{\cf{RunOneCodeRef}} (\cf{RunOneCodeCommitUtc}~UTC) \\
  Post-step code & both passes ran commit \texttt{\cf{PinPostCode}}; the archive carries \texttt{\cf{PinPostZipCode}}, the same ops plus
    one that our passes did not run \\
  Rows and reference heartbeats & the files run~3 staged, sha256 \texttt{\cf{PinRowsSha}}\ldots{} and \texttt{\cf{PinRefhbSha}}\ldots \\
  Style-rewrite example pairs & sha256 \texttt{\cf{PinExemplarsSha}}\ldots{} (in the archive; the same file in runs 1--3, first
    committed \cf{ExemplarsFixedUtc}) \\
  Run~3 output (sha256) & \texttt{\scriptsize\cf{RunThreeSubmissionShaTwoFiveSix}} \\
  Post-step pass~2 output & sha256 \texttt{\cf{PinPostTwoOutSha}}\ldots \\
  Upload file (sha256) & \texttt{\scriptsize\cf{UploadFileShaTwoFiveSix}} \\
  Host & \cf{PinHost} (\cf{RunThreeHostVcpu} vCPU, \cf{RunThreeHostMemGb}\,GB), \cf{PinOs}, \cf{PinDocker}, \cf{PinPython} \\
  Session container & image \texttt{\cf{PinImage}}, built on \texttt{\cf{PinImageBase}} \\
  Lean & the \vxcore{BenchToolchains} toolchains the problems pin, \vxcore{BenchToolchainOldest} to \vxcore{BenchToolchainNewest}, each
    project at its pinned commit (installed by the archive's \texttt{install\_envs.sh}) \\
  Models (OpenRouter) & \texttt{\cf{ModelSolOpenrouterId}}, \texttt{\cf{ModelTerraOpenrouterId}}, \texttt{\cf{ModelArgonOpenrouterId}};
    \texttt{max\_price} pins \cf{ModelSolMaxPricePin}, \cf{ModelTerraMaxPricePin}, \cf{ModelArgonMaxPricePin} US\$ per million input /
    output tokens \\
  \bottomrule
\end{longtable}
}

\section{Post-step operations and adoption counts}
\label{app:ops}
The post-step code in the archive has \cf{PostOpsInCode} ops (later versions in our open package add more):
\cf{PostOpsVOne} in version~1 (three of the closer pass's rules as ops, and the kernel-check skip, off in our passes),
\cf{PostOpsVTwoAdded} added in version~2, \cf{PostOpsVThreeAdded} in version~3, \cf{PostOpsVFourAdded} in version~4 and
\cf{PostOpsVFiveAdded} in version~5 (not in our passes; it follows the idea in the acknowledgments). Both passes ran the \cf{PostOpsRan} ops of Table~\ref{tab:ops}: pass~1
\cf{PostPassOneCompiles} compiles in \cf{PostPassOneSeconds}\,s, adopted on \cf{PostPassOneRowsAdopted} problems; pass~2
\cf{PostPassTwoCompiles} compiles in \cf{PostPassTwoSeconds}\,s, adopted on \cf{PostPassTwoRowsAdopted}. \cf{PostOpsWithSteps} ops
adopted at least one step. A step's gain is its paired reading minus the problem's previous one; summed over all steps it is
\cf{PostOpsGainSumMean} on the \vxcore{BenchRows}-problem mean, against \cf{PostGain} for the stored scores. The ops read only the
proof, the statement and the problem's own environment. Op names are the code's identifiers; a model name in an op's name names the
research session whose rule it is. One arm of the research sessions behind version~3 also received material recovered from another
entrant's public adapter, LeanZip~\citep{leanzip}; the code README does not say so. Of that arm's ops, our passes ran
\cf{LeanzipClosedOpsRan} (\texttt{argon\_funind}, \texttt{opus\_goalmerge}, \texttt{opus\_patfuse}, \texttt{sonnet\_surface},
\texttt{sol\_harvest}); \cf{LeanzipClosedOpsAdopted} adopted \cf{LeanzipClosedSteps} steps worth $+\cf{LeanzipClosedGain}$ of the
\cf{PostOpsGainSumMean} summed above.

{\scriptsize
\setlength{\tabcolsep}{3pt}
\begin{longtable}{@{}p{0.15\linewidth}cp{0.55\linewidth}rrr@{}}
  \caption{Post-step ops of the upload's passes: adopted steps, problems with a step, and the summed gain of its steps on the
    \vxcore{BenchRows}-problem mean (both passes).}\label{tab:ops}\\
  \toprule
  Op & Version & What it proposes & Steps & Problems & Gain \\
  \midrule
  \endfirsthead
  \toprule
  Op & Version & What it proposes & Steps & Problems & Gain \\
  \midrule
  \endhead
  \ClosedOpsRows
  \bottomrule
\end{longtable}
}

\section{Upload history}
\label{app:uploads}
Table~\ref{tab:uploads} lists every upload of this username to the final board. Times: the release \vxcore{ReleaseUtc}; run~1
\cf{RunOneLaunchUtc} to \cf{RunOneEndUtc}; run~2 \cf{RunTwoLaunchUtc} to \cf{RunTwoEndUtc}; run~3 \cf{RunThreeLaunchUtc} to
\cf{RunThreeEndUtc}; the post-step passes \cf{PostPassOneStartUtc} to \cf{PostPassTwoEndUtc}; the deadline \cf{FinalDeadlineUtc}
(set in AoE time; the final board was read at \cf{FinalBoardReadUtc}). Against the previous upload without the kernel-check skip (whose post-step
differed; against run~3's own output no problem is lower), the final upload keeps \cf{DiscSubThreeFourSameLines} of the
\vxcore{BenchRows} proofs unchanged, scores higher on \cf{DiscSubThreeFourHigher} problems and lower on \cf{DiscSubThreeFourLower}
(mean $+\cf{DiscSubThreeFourGain}$ on our meter). Before the release this username made \cf{WarmupPhaseUploads} uploads to the warm-up
board (\cf{WarmupPhaseFirstUtc} to \cf{WarmupPhaseLastUtc}); \cf{WarmupPhaseUncappedUploads} of them, including the last one that board
showed (\cf{WarmupBoardLastCombined}), were development results not produced under the cap from the reference proofs.

{\small
\setlength{\tabcolsep}{4pt}
\begin{longtable}{@{}p{0.17\linewidth}p{0.45\linewidth}rrrr@{}}
  \caption{Closed-track uploads to the final board. ``Ours'': the mean our meter predicted; $L$ and $H$ as graded (our upload record
    has no $L$ and $H$ for one upload).}\label{tab:uploads}\\
  \toprule
  Sent (UTC) & What & Ours & Board & $L$ & $H$ \\
  \midrule
  \endfirsthead
  \toprule
  Sent (UTC) & What & Ours & Board & $L$ & $H$ \\
  \midrule
  \endhead
  \ClosedUploadRows
  \bottomrule
\end{longtable}
}

\section{Disclosure details}
\label{app:disclosures}
The code README states what the closed package carries from our development work on the final benchmark. In its words, with internal
names translated (pass, bet and upload codes; its ``rows'' are problems here); our additions are in square brackets:

\begin{itemize}
  \item \textbf{The style rewrite's example pairs.} Its prompt carries \cf{DiscExemplarPairs} fixed example pairs (a proof before and
    after a rewrite into the scored idiom), in \texttt{harness/style\_rw\_exemplars.json} [kept out of the code tree and given to the
    organizers in the code archive; first committed \cf{ExemplarsFixedUtc}, before the release]. A pair is left out only when its
    theorem name equals the name of the problem being rewritten. \cf{DiscExemplarFinalPairs} of the \cf{DiscExemplarPairs} pairs are
    proofs of \cf{DiscExemplarTheorems} warm-up theorems that are also problems of the final benchmark, so each of those problems does
    not see its own pairs. One pair is of a [CSLib library] theorem that is a final problem under a later upstream name: the name check
    does not leave it out, so that problem's prompt shows a proof of its own theorem. [In our three runs that prompt was sent in
    \cf{DiscExemplarSelfrowPromptsSent} call, and the style rewrite changed that problem's proof in \cf{DiscExemplarSelfrowStylerwChangedRuns}
    runs.]
  \item \textbf{The end-stage passes of runs 2 and 3.} The hint pass and the closer pass (with its four variants) were developed on the
    final benchmark after its release (\vxcore{ReleaseUtc}). Their main source is our offline research: uncapped agent sessions of
    other closed models (GPT-6~Astra, Claude Opus~5.5, Fable~5.1, Gemini~4 Argon), run outside the US\$\cf{CapUsdPerProblem} cap on
    the \cf{DiscResearchRows} final-benchmark problems where our first official run was at a structural ceiling. None of their proofs
    is in any upload or package. The closer rules were learned (no model was trained) by
    comparing the research proofs and session histories with our first run's final proofs of those problems, then sized and smoke-tested on that run's \vxcore{BenchRows} final proofs. Run~3's two closer
    variants came from two later research proofs on two of the problems, compared with the closer pass's output on those final proofs.
    The hint pass's constant rule came from the research on one of the problems, and its list-minimizing rule from a per-problem
    analysis of our first run's \vxcore{BenchRows} problems, made from that run's state about \cf{DiscGrindminHoursBeforeEnd}\,h before
    it ended (pruning \texttt{grind} hints was already a known lever from our warm-up work). The dialect pass's rules were written
    before the final benchmark was released, from our warm-up work: comparisons of our pipeline's warm-up final proofs with the best
    proofs of our earlier research lines (uncapped, several closed models) on the \vxcore{WarmupRows} warm-up problems,
    \vxcore{WarmupSharedRows} of which are also final-benchmark problems. The passes hold no problem names and no proof text of any
    problem: their templates are the generic shapes of the research proofs, with the problem-specific names replaced by what Lean finds
    in the problem's own environment. The only fixed library names they write are \texttt{forall\_imp} and \texttt{congrArg} (closer
    pass), \texttt{Function.dcomp} (one variant) and one variant's \cf{DiscSimpnormLemmas} normalisation lemmas, a list widened from the
    three (\texttt{zsmul\_eq\_mul}, \texttt{ite\_mul}, \texttt{zero\_mul}) that a research proof of one of the problems added. All other
    library names they write are the environment's answers (such as the implication lemma a query finds). [Each pass is gated like the
    end stage, with a \cf{DiscPassToolS}\,s tool phase and at most \cf{DiscPassCompilesDialect} / \cf{DiscPassCompilesHint} /
    \cf{DiscPassCompilesCloser} compiles for the dialect / hint / closer pass.]
  \item \textbf{Which passes run.} We chose the passes these runs switch on from measurements on final-benchmark problems: \$0 replays
    on our first run's final proofs (in sample for the closer pass: most of its measured gain is on the research problems its rules
    came from), a paired control-vs-bets model run on \cf{DevBetsRows} final-benchmark problems in which three paid bets (tested
    together with the dialect pass) showed no evidence of gain ($+\cf{DevBetsGain}$ a problem, 95\,\% CI \cf{DevBetsCiLo} to
    $+\cf{DevBetsCiHi}$; one of the bets never fired), so the three bets were dropped (the dialect pass was kept on the \$0 replay),
    and, for one closer variant, the Arena's per-problem score of our run-1 probe upload, which showed that the official worker reads a
    final proof that keeps a search tactic as our meter does. Two other planned changes were skipped without a test.
  \item \textbf{The fixes of run~2} come from run~1's records on the final benchmark: the environment fix from the
    \cf{DiscRunOneEnveditRows} problems whose references failed, the rescore on a reference change from \cf{DiscRunTwoRescoreRows} of
    them, the no-headroom stop from one problem (the only problem of this benchmark it applies to), the idle compile slots for the brief
    from one slow problem, and the router's use of the run's references from the \vxcore{WarmupSharedRows} warm-up-named problems.
    [Run~1 made the environment edit by hand, \cf{DiscRunOneEnveditMinutes} minutes into the run, after sessions had banked bests
    against placeholder references.]
  \item \textbf{The post-step}, which our upload went through after the run: its \cf{DiscPostOpsTotal} ops (every one but the
    kernel-check skip in the default list; our upload's passes ran \cf{PostOpsRan}, Appendix~\ref{app:ops}) were developed on the final
    benchmark after its release: the \cf{PostOpsVTwoAdded} of its version~2 from a static study of our runs' proofs of its problems,
    this run's among them; those of versions~3 to~5 by research sessions and agent sessions of ours that compared our open-track and
    closed-track proofs of its problems. The ops hold no problem names and no proof text of any problem. [One arm of those research
    sessions also received LeanZip material (Appendix~\ref{app:ops}).]
  \item \textbf{\texttt{data/}} holds the \vxcore{WarmupRows} public warm-up problems and their reference heartbeats
    (\vxcore{WarmupSharedRows} of the \vxcore{WarmupRows} are also final-benchmark problems), which the code reads as reference data.
\end{itemize}

\noindent Also disclosed in Section~\ref{sec:lessons}: the kernel-check skip on \cf{DiscKskipRows} problems in two earlier uploads,
withdrawn before the final upload; the docstrings and comments in the harness code, readable by the sessions but read by none, that name library lemmas
from our warm-up work for \cf{IsoDocstringNamedRows} final problems (for \cf{IsoDocstringNamedInSeed} of them the lemma is already in
its seed proof) and give local names or unnamed hints for \cf{IsoDocstringHintRows} more; run~3 on a second host of the same type and the choice of the better whole run; the two changes
to the frozen code (a fix so that the first session sees the auto-brief's start, shipped; an opt-in mode without session containers,
unused); the hand edit of \cf{DiscRunOneEnveditRows} problems' environment files \cf{DiscRunOneEnveditMinutes} minutes into run~1; and this
username's warm-up uploads.

\section{A before/after proof example}
\label{sec:example}
Figure~\ref{fig:example} shows the Strata problem \texttt{Lambda.liftBVars\_typeCheck}. The reference proof inducts on the
expression, one arm per constructor; both finals use functional induction along \texttt{LExpr.typeCheck}, whose recursion already
splits the cases the way the goal needs, and close every case with a \texttt{simp\_all} and a \texttt{grind} call.

The routes differed. In the closed run, GPT-6.1 Sol's sessions found the functional-induction proof in the first budget step,
which ended at US\$\vxcore{ExClosedFirstSpend} of the problem's spend (\vxcore{ExClosedFirstScore} on our meter); the cheaper
model's later steps did not improve it, and the review (Gemini 4 Argon) saved a token with dot notation (\texttt{e.typeCheck} is a
single token for the lexer), for \vxcore{ExClosedFinalScore} at a total spend of US\$\vxcore{ExClosedTotalSpend}. In the open run,
the exported proof (GLM-5.3-Flash's sessions, then the end stage) kept the structural induction (\vxcore{ExOpenExportLines} lines,
\vxcore{ExOpenExportScore}); the post-step's functional-induction op rewrote it in pass \vxcore{ExOpenFunindPass}
(\vxcore{ExOpenFunindScore}), and smaller edits up to pass \vxcore{ExOpenPostLastPass} reached \vxcore{ExOpenFinalScore}. That op is
one of those designed after the release by comparing our open and closed-track proofs of the final problems
(Section~\ref{sec:endstage}).

The open final is \vxcore{ExTokensDiff} tokens shorter, the closed one \vxcore{ExHbDiff} heartbeats cheaper. Here a token is worth
$B/T=\vxcore{ExRate}$ heartbeats, so the heartbeat saving is worth about \vxcore{ExHbDiffInTokens} tokens and the closed proof scores
\vxcore{ExScoreDiff} points higher. Its closers are restricted (\texttt{simp\_all!~only}, \texttt{grind only} with E-matching
annotations) where the open proof's \texttt{simp\_all} uses the default simp set; we did not profile the gap.

\begin{figure}[t]
  \centering
  \lstset{basicstyle=\ttfamily\scriptsize}
  \begin{minipage}{\linewidth}
    \small
    \textbf{Statement} (shared; each proof below follows its \texttt{:= by})
\begin{lstlisting}
theorem liftBVars_typeCheck
    (e : LExpr T.mono)
    {Δ₁ Δ_insert Δ_outer : List LMonoTy}
    : LExpr.typeCheck (Δ₁ ++ Δ_insert ++ Δ_outer) (LExpr.liftBVars Δ_insert.length e Δ₁.length)
      = LExpr.typeCheck (Δ₁ ++ Δ_outer) e := by
\end{lstlisting}
    \textbf{Reference proof}: \vxcore{ExSeedLines} lines, \vxcore{ExSeedTokens} tokens, \vxcore{ExSeedHb} heartbeats, score
    \vxcore{SeedScore}
\begin{lstlisting}
  induction e generalizing Δ₁ with
  | const => simp [LExpr.liftBVars, LExpr.typeCheck]
  | op m o ty => cases ty <;> simp [LExpr.liftBVars, LExpr.typeCheck]
  | fvar m name ty => cases ty <;> simp [LExpr.liftBVars, LExpr.typeCheck]
  | bvar m i =>
    simp only [LExpr.liftBVars, LExpr.typeCheck]
    split
  -- (*\textit{\vxcore{ExSeedLinesOmitted} more lines}*)
\end{lstlisting}
    \textcolor{vxOpen}{\rule{0.9ex}{0.9ex}}~\textbf{Open upload}: \vxcore{ExOpenFinalTokens} tokens, \vxcore{ExOpenFinalHb}
    heartbeats, score \vxcore{ExOpenFinalScore}
\begin{lstlisting}
  fun_induction LExpr.typeCheck Δ₁ e
  all_goals simp_all -contextual [LExpr.liftBVars, LExpr.typeCheck]
  grind only [LExpr.typeCheck, List.getElem?_append]
\end{lstlisting}
    \textcolor{vxClosed}{\rule{0.9ex}{0.9ex}}~\textbf{Closed upload}: \vxcore{ExClosedFinalTokens} tokens, \vxcore{ExClosedFinalHb}
    heartbeats, score \vxcore{ExClosedFinalScore}
\begin{lstlisting}
  fun_induction e.typeCheck Δ₁ <;> simp_all! only [List.cons_append]
  grind only [= LExpr.typeCheck.eq_6, = List.getElem?_append, = List.length_append]
\end{lstlisting}
  \end{minipage}
  \caption{Before and after on \texttt{Lambda.liftBVars\_typeCheck} (Strata; toolchains \vxcore{ExPins}). Tokens by our copy of
    the Arena's lexer, heartbeats on the first-listed toolchain, scores on our meter. Both finals compile on every listed toolchain
    ($Z=\vxcore{FullZ}$).}
  \label{fig:example}
\end{figure}

\noindent The functional-induction op (\texttt{argon\_funind}) comes from the research arm that also received LeanZip material; our closed
passes ran it with no adoption (Appendix~\ref{app:ops}), and our closed proof of this problem comes from Sol's sessions.

\end{document}